\documentclass[11pt,final, sort&compress]{elsarticle}

\usepackage{amssymb}
\usepackage{amsmath}
\usepackage{graphicx}
\usepackage{tabularx}
\usepackage{cite}
\usepackage{color}
\usepackage{fullpage}
\usepackage{subfigure}
\usepackage{subfigmat}
\usepackage{etoolbox}
\usepackage{url}
\usepackage{float}
\usepackage{bm}
\usepackage[dvipsnames]{xcolor}
\usepackage{siunitx} 

\journal{International Journal of Multiphase Flow}

\begin{document}
\begin{frontmatter}

\title{
Investigation of condensation shocks and re-entrant jet dynamics in a cavitating nozzle flow by Large-Eddy Simulation
}

\author{Theresa Trummler}\corref{cor}
\cortext[cor]{Corresponding author}
\ead{theresa.trummler@tum.de}
\author{Steffen J. Schmidt}
\author{Nikolaus A. Adams}

\address{
    Chair of Aerodynamics and Fluid Mechanics, Technical University of Munich \\ 
    Boltzmannstr. 15, 85748 Garching bei M\"unchen, Germany}

\begin{abstract}

    Cloud cavitation is related to an intrinsic instability where clouds are shed periodically. The shedding process is initiated either by the motion of a liquid re-entrant jet or a condensation shock. Cloud cavitation in nozzles interacts with the flow field in the nozzle, the mass flow and the spray break-up, and causes erosion damage. For nozzle geometries cloud shedding and the associated processes have not yet been studied in detail. 

    In this paper, we investigate the process of cloud cavitation shedding, the re-entrant jet and condensation shocks in a scaled-up generic step nozzle with injection into gas using implicit Large-Eddy Simulations (LES). For modeling of the cavitating liquid we employ a barotropic equilibrium cavitation model, embedded in a homogeneous multi-component mixture model. Full compressibility of all components is taken into account to resolve the effects of collapsing vapor structures.

    We carry out simulations of two operating points exhibiting different cavitation regimes. The time-resolved, three-dimensional simulation results cover several shedding cycles and provide deeper insight into the flow field. Our results show that at lower cavitation numbers, shedding is initiated by condensation shocks, which has not yet been reported for nozzle flows with a constant cross-section. We analyze the cavitation dynamics and the shedding cycles of both operating points. Based on our observations we propose modifications to established schematics of the cloud shedding process. Additionally, we analyze the near-wall upstream flow in and underneath the vapor sheet and possible driving mechanism behind the formation of the re-entrant jet.

\end{abstract}


\end{frontmatter}
\section{Introduction}
\label{sec:Intro}

Cavitation phenomena in nozzles play a crucial role for various hydraulic systems such as dump throttles of liquid propellant rocket engines, orifices of hydraulic jacks and injector components. For the latter, cavitation can affect mass flow~\citep{nurick1976orifice, payri2004influence, Trummler:2018AAS}, spray break up~\citep{reitz1982mechanism, Sou:2007jd, Orley:2015kt} and resilience against cavitation erosion~\citep{asi2006failure}. Understanding of cavitation dynamics and of underlying mechanisms of cavitation in injector nozzles is essential.

Cloud cavitation is a form of partial cavitation where vapor clouds periodically shed from the main cavity~\citep{reisman1998observations, laberteaux2001partial}. This form of cavitation occurs in flows around bodies, such as hydrofoils~\citep{Le:1993partial, Kubota:1989unsteady, de1997sheet}, and in internal flows, e.g. Venturi nozzles~\citep{rudolf2014characterization,Hayashi:2014ik}, converging-diverging ducts with a rectangular cross-section (wedges)~\citep{Stutz:1997ui, ganesh2016bubbly}, and other types of orifices and nozzles~\citep{Stanley:2014id, Stanley:2011gr, Sato:2002vv, Saito:2003us, Sugimoto:2009, Sou:2007jd, He:2016bx}. The periodic cloud shedding process is initiated by the upstream motion of a disturbance, which can be either a thin liquid film underneath the cavity (\textit{re-entrant jet}) or a bubbly \textit{condensation shock}. 

A re-entrant jet characterizes the motion of a liquid jet underneath the fixed cavity in upstream direction and is generally considered to initiate the shedding process. \citet{Le:1993partial} proposed a schematic description for the periodic cloud shedding process on hydrofoils, which \citet{Stanley:2014id} adapted for internal nozzle flows. After the collapse of a shed cloud, a re-entrant jet develops at the end of the cavity and travels upstream underneath the cavity. When the re-entrant jet has covered the entire cavity and reached the inception point of the cavity, the cavity is shed and a new one begins to form~\citep{Le:1993partial, wade1966experimental}. The shed cloud convects downstream with the flow, in a rolling motion~\citep{Le:1993partial, Kubota:1989unsteady}. In the region of increased pressure downstream the shed cloud collapses, emitting an intense shock wave leading to pressure peaks orders of magnitude larger than the pressure in the mean flow~\citep{reisman1996pressure, reisman1998observations}, which can cause cavitation erosion~\citep{gopalan2000flow,petkovvsek2013simultaneous}. Moreover, these cavitation induced pressure fluctuations \citep{Leroux:2004jl,ji2015large} are also considered to contribute to the formation of the re-entrant jet, as discussed below. 

Although many experimental and numerical studies have investigated re-entrant jets and their decisive role in cloud shedding (e.g.~\citet{kawanami1997mechanism, lush1986high, gopalan2000flow, Furness:1975wq}), the driving mechanism behind the formation of the jet has not yet been clarified. One theory is that the re-entrant jet is formed due to the stagnation point behind the cavity (at the closure), where the flow surrounding the cavity impinges on the wall. Another suggestion is that the pressure peaks due to the collapse of shed clouds promote the motion of the re-entrant jet~\citep{Leroux:2004jl,Leroux:2005ii}. \citet{CoutierDelgosha:2007dp}, however, found no general relation between the shock waves after cloud collapse and the re-entrant jet motion. \citet{callenaere:2001} consider the cavity thickness in relation to the re-entrant jet thickness and the negative pressure gradient to be the two most relevant parameters for the instability of the re-entrant jet. 

Several experimental investigations have assessed the velocity of re-entrant jets. For the cavitating flow on a hydrofoil, \citet{Pham:1999tp} measured that the re-entrant jet velocity at the end of the cavity is about the free stream velocity and decreases upstream. Decreasing upstream jet velocity was also observed by \citet{Sakoda:2001tn}. \citet{Le:1993partial} found that the velocity of perturbations traveling upstream is close to the magnitude of the free stream flow. \citet{Stanley:2014id} tracked bubbles in the liquid film underneath the cavity and obtained velocities of 20 - 30\% of the free-stream velocity, which is a smaller magnitude than reported in previous findings. 

In addition to re-entrant jets, condensation shocks can also initiate periodic cloud shedding. The occurrence of condensation shocks was predicted back in 1964 by \citet{jakobsen1964mechanism}. Later, condensation shocks on hydrofoils have been studied experimentally by e.g.~\citet{reisman1998observations, Arndt:2001tx}. Converging-diverging geometries have been investigated experimentally by~\citet{ganesh2016bubbly, Jahangir:2018id, Wu:2017cda, Wang:2017hf} and recently by compressible numerical simulations~\citep{budich2018jfm}. For shock initiated shedding, one observes an upstream moving bubbly shock rather than a re-entrant jet. The formation of bubbly shocks is more likely when the cavitation number decreases, and for increased void fraction in the cavity~\citep{ganesh2016bubbly}. We are not aware of publications addressing the occurrence of condensation shocks in nozzles with constant cross-sections. However, the observations by \citet{Stanley:2014id} may be explained by a condensation shock mechanism. 

The flow field within realistic geometries, such as injector components or control valves, is difficult to access due to limited optical accessibility, small dimensions and complex geometries as well as high velocities and pressures. Therefore, experimental investigations are mostly performed with upscaled transparent nozzles. For the analysis of cavitation dynamics in orifices and control valves, experiments are carried out with injection into liquid (submerged injection)~\citep{ Sato:2002vv, Saito:2003us,Sugimoto:2009}. Cavitation in injector components and the interaction of cavitation with jet break up was investigated by~\citet{ Stanley:2014id, Stanley:2011gr,Sou:2007jd,Sou:2014hja, He:2016bx} with injection into gas. Time-resolved numerical simulations are not limited by the above mentioned constraints and provide three dimensional flow field data to assess the processes in realistic geometries and under real conditions. LES can capture the interaction of cavitation and turbulence, as shown by \citet{Egerer:2014wu}. A large number of numerical investigations~\citep{Orley:2015kt, Trummler:2018AAS,Koukouvinis:2016boa, Edelbauer:2017fx, Bicer:2015cp, Bicer:2017ia} have used the above mentioned reference experiments~\citep{Sou:2007jd,Sou:2014hja} to validate their modeling approaches and thereby provided deeper insights into the underlying physical processes. Successfully validated models were applied to realistic injector geometries~\citep{Orley:2016db, Koukouvinis:2016boa}. 

Based on the above mentioned studies, two cavitation regimes can be identified. The first regime involves developing cavitation, where cavitation is initiated in the shear layer close to the nozzle inlet and the cavity length reaches about half of the nozzle length~\citep{Sou:2007jd, Stanley:2011gr}. In the separated shear layer at the nozzle inlet, cavitating spanwise vortices form, coalesce, and pair in a rotational motion to larger clouds~\citep{Sato:2002vv, Saito:2003us}. In the second regime, known as super cavitation~\citep{Chaves:1995vt}, inertia-driven cavitation usually produces a cavity length larger than 80\%. Clouds detach from the end of the sheet and collapse close to the nozzle outlet. \citet{Stanley:2014id} found that transition between these two regimes is characterized by a cavity-length jump and aperiodic shedding. The cavity-length jump can also be seen in other experimental measurements~\citep{Sou:2007jd, Saito:2003us}. In this strongly cavitating regime, the shed clouds can be transported to the nozzle outlet and generate a pressure gradient from the low-pressure vapor region to the outflow, which results in gas being ingested by the nozzle~\citep{Orley:2015kt, Trummler:2018AAS}. Further increase of the cavitation length can result in a complete flow detachment from the nozzle wall and the so-called hydraulic flip~\citep{Sou:2007jd, Stanley:2011gr}. 

This paper provides a comprehensive analysis of cavitation dynamics in internal nozzle flows with constant cross-section and discharge into ambient gas. We perform compressible implicit LES at two operating points following a reference experiment~\citep{Sou:2014hja, Bicer:2015cp}. We analyze cavitation dynamics and the initiating shedding mechanisms. Based on our results, we propose modifications to schematics of the cloud shedding process~\citep{Stanley:2014id, Le:1993partial}. Moreover, we evaluate the near wall flow field and investigate mechanisms for re-entrant jet formation. 

We apply a model proposed by \citet{Orley:2015kt}, which is an extension of the equilibrium cavitation model of \citet{Schnerr:2008jja} and \citet{schmidt2014assessment}. The thermodynamic model is embedded into a density-based, fully compressible flow solver with an implicit higher-order LES approach for compact stencils~\citep{Egerer:2016it}.

The paper is structured as follows. In section~\ref{sec:MathePhysModel} the governing equations, the thermodynamic models and the numerical approach are introduced. Section~\ref{sec:NumSetup} presents the computational setup. In section~\ref{sec:Results}, we compare our computational results with experimental data, analyze the cavitation dynamics and the shedding mechanisms in detail and relate them to the flow field. 

\section{Mathematical and physical model}
\label{sec:MathePhysModel}

\subsection{Governing Equations}

We solve the fully compressible Navier-Stokes equations in conservative form 
    \begin{equation}
        \partial_{t}\boldsymbol{U}+\nabla \cdot [ \boldsymbol{C}(\boldsymbol{U})+\boldsymbol{S}(\boldsymbol{U})]=0\,. 
        \label{eq:NS}
    \end{equation}
The state vector $\boldsymbol{U}=[\rho , \, \rho u_{1}, \, \rho u_{2} , \, \rho u_{3} , \, \rho \xi ]$ is composed of the conserved variables density $\rho$, momentum $\rho u_{i}$ and a non-condensable gas content $\rho \xi$, where $\xi$ represents the gas mass fraction. Inclusion of non-condensable gas into the model allows for simulating an injection into air as well as the prediction of gas entrainment. Due to the barotropic modeling of the cavitating liquid, combined with an isothermal model of the gas phase, an energy equation is not needed. The convective fluxes $\boldsymbol{C}(\boldsymbol{U})$ across finite-volume cell faces are 
    \begin{equation}
        \boldsymbol{C}_{i}(\boldsymbol{U})=u_{i}\boldsymbol{U}\;
        \label{eq:C}
    \end{equation}
and the cell-face stresses $\boldsymbol{S}(\boldsymbol{U})$ are
    \begin{equation}
        \boldsymbol{S}_{i}(\boldsymbol{U})=[0,\delta_{i1}p-\tau_{i1}, \delta_{i2}p-\tau_{i2}, \delta_{i3}p-\tau_{i3},  0 ] \,,
        \label{eq:S}
    \end{equation}
where $p$ is the static pressure, $\delta_{ij}$ the Kronecker-Delta and $\boldsymbol{\tau}$ the viscous stress tensor 
    \begin{equation}
        \tau_{ij}=\mu (\partial_{j}u_{i}+\partial_{i}u_{j}+\frac{2}{3}\delta_{ij}\partial_{k}u_{k}) \,,
        \label{eq:tau}
    \end{equation}
with a dynamic mixture viscosity $\mu$, which depends on the fluid composition in the cell as discussed below. 

\subsection{Thermodynamic model}

For the modeling of two-phase flows, homogeneous mixture models~\citep{Kubota:1992eo} describing the multiphase mixture with a compressible effective fluid in a finite volume cell have been established and are widely used for different cavitation regimes, such as re-entrant jet dominated cavitation over a wedge~\citep{Gnanaskandan:2016bi}, condensation shock governed cavitation over a wedge~\citep{budich2018jfm} and cavitation erosion prediction~\citep{mihatsch2015cavitation,beban2017numerical}. 

\citet{Schnerr:2008jja} formulated a cavitation model based on the assumption that the liquid and gas phases of a cavitating liquid are in thermal and mechanical equilibrium. The thermodynamic equilibrium assumption implies a physically consistent sub-grid model for vapor structures~\citep{schmidt2014assessment}, making such a model particularly suitable for LES. Furthermore, this model is parameter-free and able to treat appropriately various cavitation regimes~\citep{Egerer:2014wu}. The model has been extensively validated~\citep{Schmidt:2015wa, schmidt2014assessment, mihatsch2015cavitation,beban2017numerical,budich2018jfm}. \citet{Orley:2015kt} extended the single-fluid cavitation model by a component of non-condensable gas to a multi-component cavitation model and performed LES of the primary break-up phase of cavitating water jets injected into ambient gas. 

In this study, we use the multi-component cavitation model of \citet{Orley:2015kt}. All components are represented by an effective fluid considering a cell-averaged equilibrium pressure $p$, equilibrium velocity $\boldsymbol{u}$ and Temperature $\vartheta$. Sub-cell phase interfaces are not reconstructed and surface tension is neglected. We consider the components pure liquid, liquid-vapor-mixture, and non-condensable gas, which are denoted by the indices $L$, $M$, and $G$, respectively. The volume fraction $\beta$ of each component $\Phi = \{ L, M, G\}$ within a control volume $V$ is
    \begin{equation}
        \beta_{\Phi}=\frac{V_{\Phi}}{V} \quad \mathrm{with} \quad \sum_{\Phi} \beta_{\Phi}=1,
        \label{eq:beta}
    \end{equation} 
and the mass fraction $\xi$ of the mass $m$ of this control volume is
\begin{equation}
        \xi_{\Phi}=\frac{m_{\Phi}}{m} \quad \mathrm{with} \quad \sum_{\Phi} \xi_{\Phi}=1. 
        \label{eq:xi}
    \end{equation} 
The density of each component is 
\begin{equation}
        \rho_{\Phi}=\frac{m_{\Phi}}{V_{\Phi}}
        \label{eq:xi}
    \end{equation} 
and the mixture density is 
    \begin{equation}
        \rho=\sum_{\Phi} \beta_{\Phi} \rho_{\Phi} \,. 
        \label{eq:rho}
    \end{equation}
Combining pure liquid and liquid-vapor-mixture to $LM$, the mixture density is given by
    \begin{equation}
        \rho=\beta_{LM} \rho_{LM}+ \beta_{G} \rho_{G}= (1-\beta_{G}) \rho_{LM}+ \beta_{G} \rho_{G}\,. 
        \label{eq:rho2}
    \end{equation}     
In the following, we express $\rho_{LM}$ and $\rho_{G}$ by barotropic relations and combine them to a coupled equation of state (EOS)~\citep{Trummler:2018AAS, Orley:2015kt}. The volume fraction of gas $\beta_{G}$ in Eq.\eqref{eq:rho2} can be expressed by the transported mass fraction of gas $\xi_{G}$ and the mixture density as 
    \begin{equation}
        \beta_{G}=\frac{V_{G}}{V}=\frac{m_{G}}{\rho_{G}\,V}=\frac{m_{G}/m}{\rho_{G}\,V/m}=\xi_{G}\frac{\rho}{\rho_{G}}. 
        \label{eq:rho}
    \end{equation}     

The equation of state for the pure liquid and the two-phase region is derived from the definition of the isentropic (entropy $s=\mathrm{const.}$) speed of sound $c$
    \begin{equation}
        c=\sqrt{\frac{\partial p}{\partial \rho}\bigg|_{s=\mathrm{const.}}}\;\;.
        \label{eq:c0}
    \end{equation}
Assuming an approximately constant speed of sound in the liquid region and also in the two-phase region, one can integrate Eq.~\eqref{eq:c0} for each region starting from saturation density $\rho_\mathrm{sat,liq}$ to obtain a linearized relation between density and pressure valid in both regions
    \begin{equation}
        \rho_{LM}=\rho_\mathrm{sat,liq}+\frac{1}{c^2}(p-p_\mathrm{sat}) \;\;\mathrm{with}\;\;c=
        \left\{
            \begin{array} {lcl}c_{L}  & \mathrm{if} & p\geq p_\mathrm{sat}\\
            c_{M} & \mathrm{if} & p < p_\mathrm{sat}
        \end{array}\right.
        .
        \label{eqn:eos_cav_liq}
    \end{equation}
For pure liquid ($p \geq p_\mathrm{sat}$) the speed of sound in the liquid $c = c_L$ is used and in the two-phase regions we apply the speed of sound of a liquid-vapor-mixture $c = c_{M}$. For water at reference temperature $\vartheta = 293.15\,\si{K}$ we set $p_\mathrm{sat}=2340\,\si{Pa}$, $\rho_\mathrm{sat,liq}=998.16\,\si{kg/m^3}$, $c_{L}= 1482.35\,\si{m/s}$ and $c_{M}=1\,\si{m/s}$~\citep{Trummler:2018AAS, Orley:2015kt}. The vapor content $\alpha$ in a computational cell is calculated from the density of the liquid vapor mixture $\rho_{LM}$ as 
    \begin{equation}
        \alpha=\frac{\rho_\mathrm{sat,liq}-\rho_{LM}}{\rho_\mathrm{sat,liq}-\rho_{\mathrm{sat,vap}}},\; \;\; \mathrm{if} \;\;\rho_{LM} < \rho_\mathrm{sat,liq} ,
    \end{equation}
where $\rho_{\mathrm{sat,vap}}$ denotes the saturation density of the vapor and is $\rho_{\mathrm{sat,vap}}= 17.2\cdot 10^{-3}\,\si{kg/m^3}$ at the reference conditions. 

The non-condensable gas-phase is modeled as an isothermal ideal gas 
    \begin{equation}
        \rho_{G}= \frac{p}{\mathrm{R}\,\vartheta}, 
        \label{eq:ideale_gasgl}
    \end{equation}
with the specific gas constant $\mathrm{R}= 287.06\,\si{J/(kg\cdot K)}$ for air and the reference temperature $\vartheta= 293.15\,\si{K}$. 

We model the dynamic viscosity of the liquid-vapor mixture $\mu_{LV}$ as proposed by \citet{Beattie:1982ut} with 
    \begin{equation}
        \mu_{LM}=\alpha\cdot\mu_{V}+(1-\alpha)(1+\frac{5}{2}\alpha)\cdot\mu_{L}. 
        \label{eq:muLiqMix}
    \end{equation}
$\mu_{V}$, $\mu_{L}$ denote the viscosity of the vapor and liquid, respectively. For a small amount of vapor the viscosity corresponds to vapor phase consisting of small vapor bubbles which behave as solid particles. The viscosity of the mixture fluid considering also the gas component is then approximated by a linear blending with the volume fractions~\citep{Trummler:2018AAS}
    \begin{equation}
        \mu=\beta_{G}\cdot\mu_{G} + (1-\beta_{G})\cdot\mu_{LM} \,.
        \label{eq:muLiqMix}
    \end{equation}
The following values for the viscosities are used: $\mu_{L} = 1.002\cdot 10^{-3}\,\si{Pa\cdot s}$ , $\mu_{V} = 9.272\cdot 10^{-6}\,\si{Pa\cdot s}$ and $\mu_{G} = 1.837\cdot 10^{-5}\,\si{Pa\cdot s}$\,.

\subsection{Numerical approach}

With LES the smallest scales are not resolved on the computational grid and therefore effects of these unresolved sub-grid-scales (SGS) have to be modeled. We employ an implicit LES approach based on the Adaptive Local Deconvolution (ALDM) method by \citet{adams2004implicit,hickel2006adaptive, hickel2014subgrid}. For implicit LES, the truncation error of the discretization serves as a sub-grid-scale model for turbulence. In this work, the implicit LES approach for compact stencils recently proposed by~\citet{Egerer:2016it} is used, which is specially designed for compressible, cavitating flows. Additionally, an extension for the additional gas phase~\citep{Trummler:2018AAS} for multiphase application is included.

Shock waves and pseudo phase boundaries are detected by sensor functionals. In the detected non-smooth regions, the velocity components are reconstructed using a third order Total-Variation-Diminishing (TVD) slope limiter by \citet{Koren:1993}, and the thermodynamic quantities $\rho$, $p$ are reconstructed using the second order minmod slope limiter~\citep{Roe:1986}. In smooth regions, a high order central discretization scheme with regularization terms is applied. The extensions introduced for the additional gas phase~\citep{Trummler:2018AAS} ensure a thermodynamically consistent coupling of the dependent variables pressure, density and gas mass fraction and thus prevent unphysical pressure fluctuations. For this purpose, a suitable reconstruction method was implemented to obtain thermodynamically consistent fluxes, especially with regard to the additional flux for the gas phase. 

Time integration is performed with an explicit second-order, 4-step low-storage Runge-Kutta method. 

\section{Numerical setup}
\label{sec:NumSetup}

\begin{figure}
\centering
    \subfigure[]{\includegraphics[width=9.4cm]{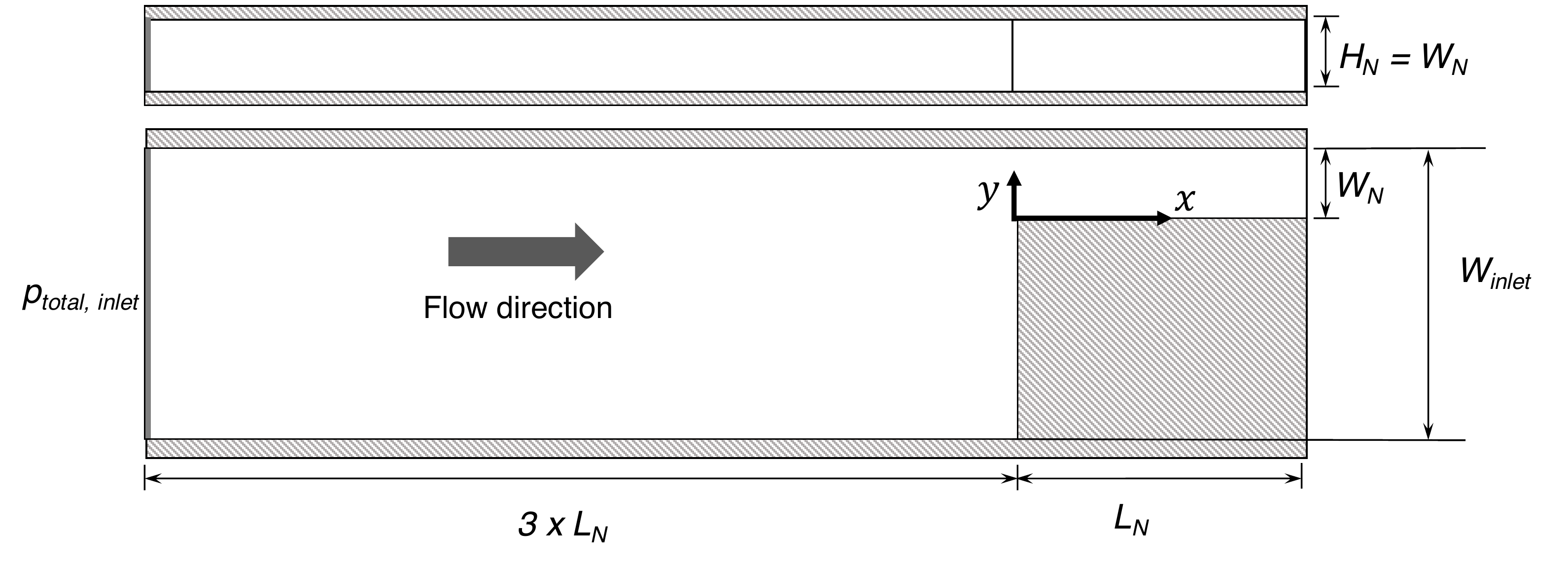}}
    \subfigure[]{\includegraphics[width=7.8cm]{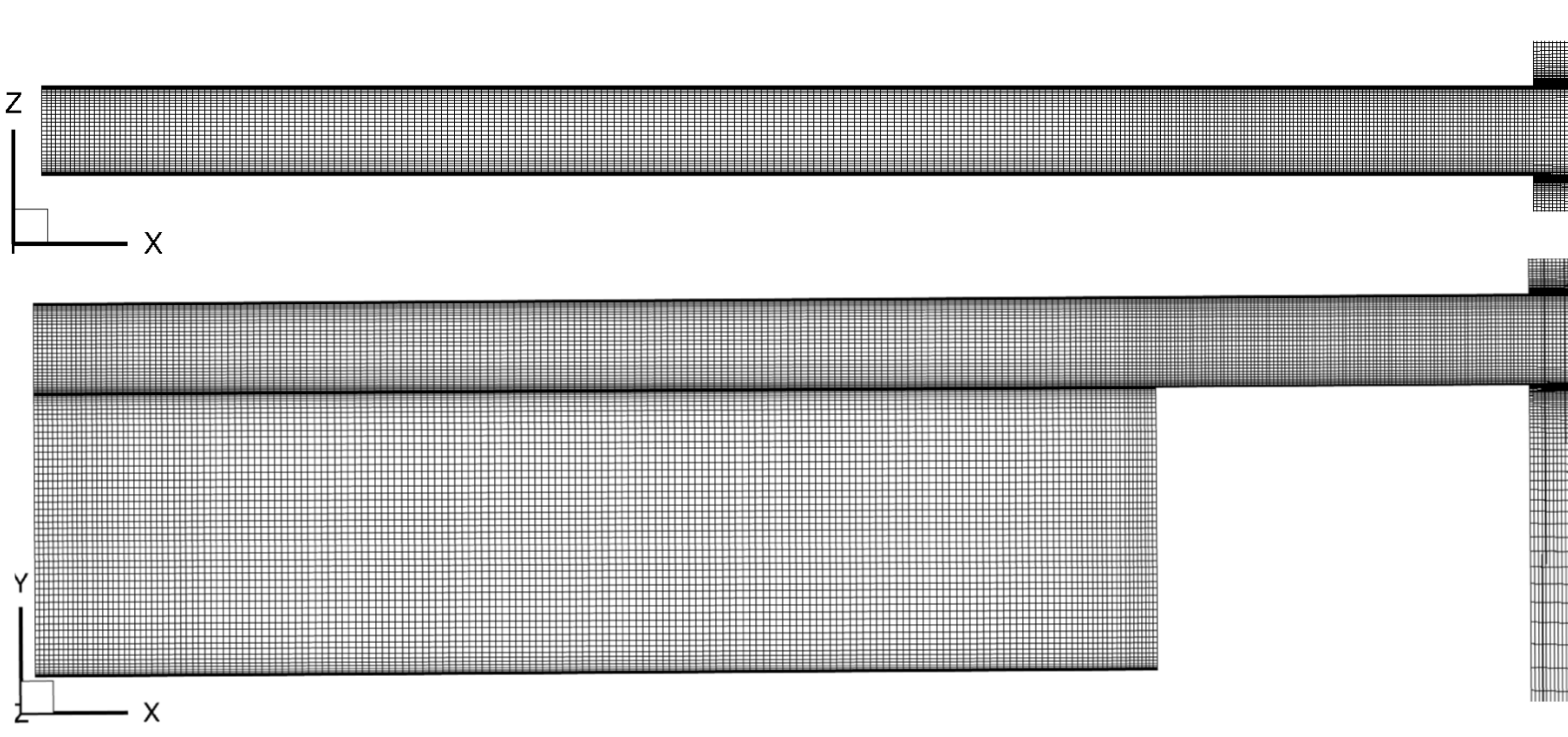}}
    \caption{
        Numerical setup (a) Sketch of the numerical setup; (b) Slices through the grid on the finest level with every fourth grid line shown. At the nozzle outlet a large domain is added ($25\,L_{N}\,\mathrm{x}\,75\,W_{N}\,\mathrm{x}\,75\,W_{N}$) and the outflow region is initialized with gas.}
\label{fig:setup}
\end{figure}%

The numerical setup follows that of an earlier investigation~\citep{Trummler:2018AAS}. The configuration is adopted from a reference experiment~\citep{Sou:2014hja, Bicer:2015cp} where water is discharged through a transparent step nozzle into ambient air. Different cavitation regimes are obtained by adjusting the inlet pressure and thus the mean velocity in the nozzle. We carry out simulations of two representative operating points, which are listed in Table~\ref{tab:overview} and explained in more detail in the following section. For both cases, the mean velocity in the nozzle is approximately $\bar{u}\approx 15\,\si{m/s}$. Fig.~\ref{fig:setup} depicts the numerical setup. The geometric dimensions are taken from the experiment and have the following values: nozzle length $L_{N}=8\,\si{mm}$, nozzle width and height (quadratic cross-section) $W_{N}=H_{N}=1.94\,\si{mm}$ and width of the inflow region $W_\mathrm{Inlet}=8\,\si{mm}$ with the same height as the nozzle $H_N$, see Fig.~\ref{fig:setup}. In the following, the nozzle inlet edge is referred to as the origin of an underlying Cartesian coordinate system. For the simulations a large domain extension is added to the nozzle outlet ($25\,L_{N}\,\mathrm{x}\,75\,W_{N}\,\mathrm{x}\,75\,W_{N}$) to prevent field boundary conditions from affecting the solution. The length of the inflow region was chosen to be three times the nozzle length and at the inlet a total pressure is prescribed, the values of which are provided in Table~\ref{tab:overview}. The outlet is defined as a pressure outlet with $p_\mathrm{out}=1 \cdot 10^{5}\;\si{Pa}$, the outlet region is initialized with gas, and the no-slip walls are isothermal.

The simulation domain is discretized on a block-based, structured grid. To reduce computational costs, we employ a grid sequencing strategy. For each operating point, we let the flow field first develop on a coarse grid and then refine the grid over several refinement steps. For a detailed description of this method including a convergence study for a comparable configuration we refer to \citet{Orley:2015kt}. Here, we use five different grid levels, where the coarsest grid contains 2.4 million cells and the finest 51.5 million cells. The results presented here are obtained on the finest grid level. On the finest grid, the smallest cell size in wall-normal direction is $2.5\,\si{\mu m}$ and the largest cell-size in the nozzle is $30.5\,\si{\mu m}$. As can be seen in Fig.~\ref{fig:setup}, the grid is refined near the nozzle walls and around the sharp edge at the nozzle inlet and outlet. Previous investigations of the same configuration~\citep{Koukouvinis:2016boa} and for comparable configurations~\citep{Orley:2015kt} have established grid convergence for the stated resolution. 

For wall-bounded turbulent flows, the dimensionless wall normal resolution $y^{+}$ is a parameter to evaluate the grid resolution. The analysis of $y^{+}$ for this configuration can be found in \citet{Trummler:2018AAS}: $y^{+} \approx 1$ within the nozzle, i.e. the near-wall cell is within the viscous sublayer.

The high grid resolution and the fully compressible modeling approach result in a time-step of $0.7\,\si{ns}$ on the fine grid with a Courant-Friedrichs-Levy number (CFL) of 1.4. The total simulation time on the finest grid was about $8\,\si{ms}$, for a flow field evolution of $4\,\si{ms}$.

\section{Results}
\label{sec:Results}

\subsection{Overview and comparison with experimental data}
\label{subsec:overview} 

The investigated configuration is a step nozzle with quadratic cross-section ($W_{N}\times W_{N}$) in which cavitation occurs due to the one-sided constriction and the velocity increase in the detached shear layer. The Reynolds number is 
    \begin{equation}
        Re=\frac{\rho \,\bar{u}\,W_{N} }{\mu},
        \label{eq:Re}
    \end{equation} 
where $\rho$ denotes the density of the liquid, $\bar{u}$ the mean velocity in the nozzle, $W_N$ the nozzle width and $\mu$ the dynamic viscosity. The cavitation number $\sigma$ is defined as
    \begin{equation}
        \sigma=\frac{p_\mathrm{out}-p_\mathrm{sat}}{0.5\,\rho\,\bar{u}^2},
        \label{eq:sigma}
    \end{equation} 
where $p_\mathrm{out}$ is the pressure at the outlet and $p_\mathrm{sat}$ the saturation pressure of the liquid. 

\begin{table}
\caption{Overview of the operating points investigated.}
\centering\begin{tabular}{|cccl|}
\hline
    $\boldsymbol{\sigma}$        \textbf{[-]} & 
    $\boldsymbol{\bar{u}}$       \textbf{[m/s]}  &  
    $\boldsymbol{p_\mathrm{ total, inlet}}$\textbf{[$10^{5}$Pa]} & 
    \textbf{Cavitation regime} \\
    \hline
    1.19  & 12.8  &  2.37   & developing    \\ 
    0.84  & 15.2  &  3.03   & super cavitation    \\ 
    \hline
    \end{tabular}%
  \label{tab:overview}%
\end{table}

\begin{figure}
    \centering
    \subfigure[]{\includegraphics[height=4.5cm]{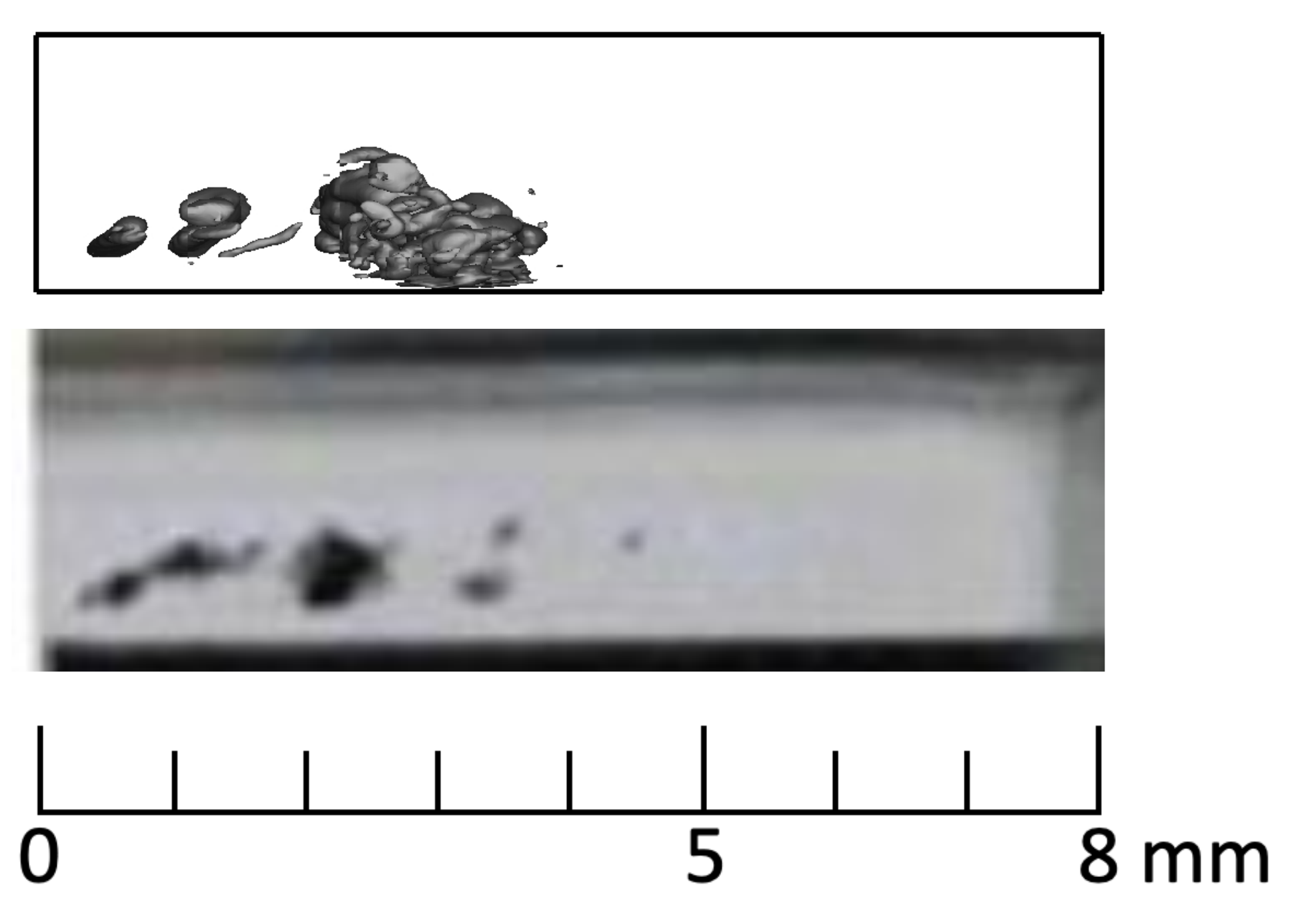}}
    \subfigure[]{\includegraphics[height=4.5cm]{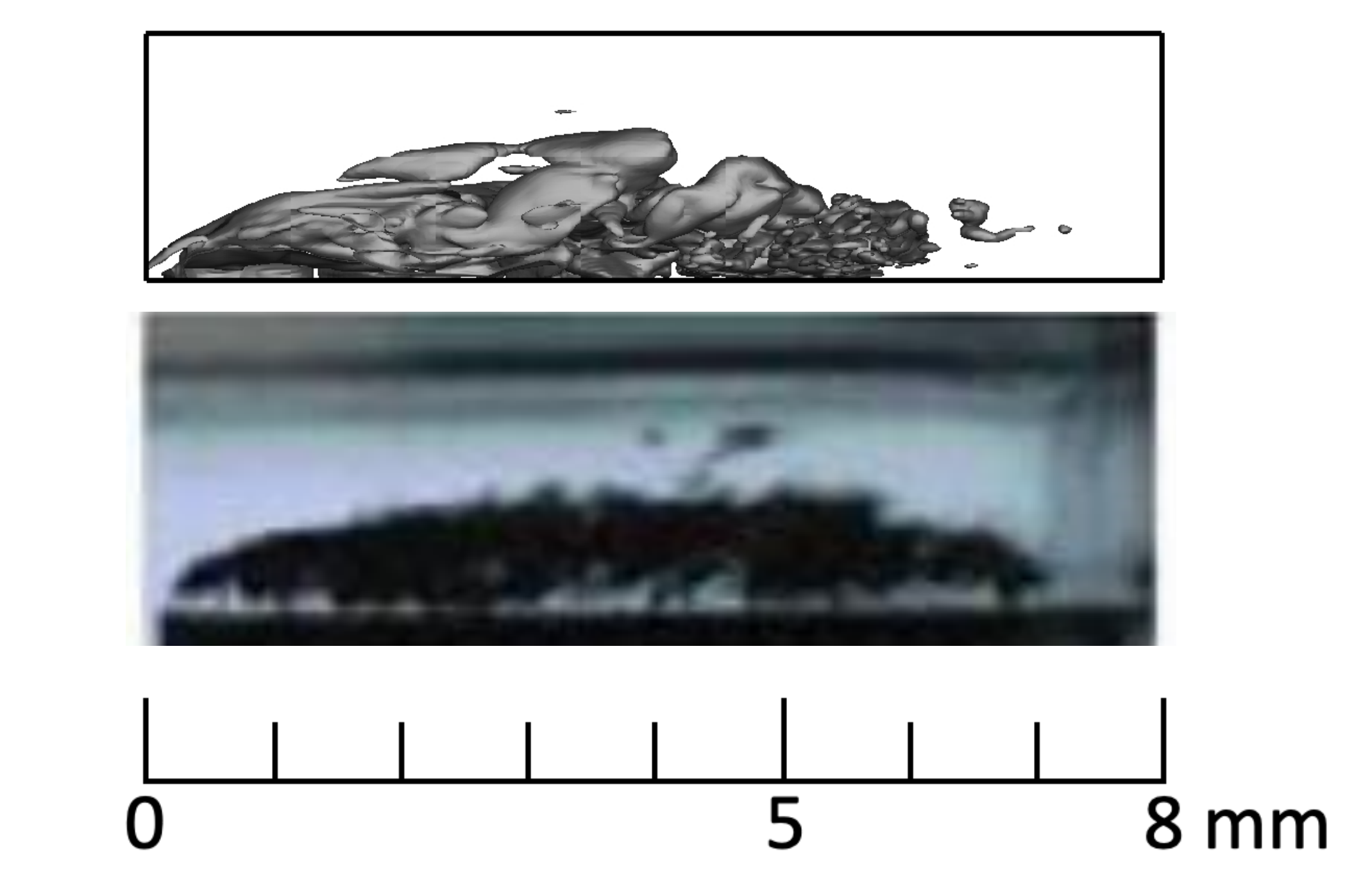}}
    \caption{
        Comparison of the experimental data \citep{Sou:2014hja} and the simulation results (Isosurface: vapor ($\alpha = 0.1$ )) (a) $\;\sigma= 1.19 $ (b)$\;\sigma= 0.84^{*}$ ($\sigma_{Experiment}= 0.82$) (experimental data is reprinted from \citep{Sou:2014hja}). Flow direction is from left to right. 
        }
\label{fig:comp_exp_sim}
\end{figure}

We have carried out simulations of two representative operating points at $\sigma = 1.19$, $Re\approx25,000$ and $\sigma = 0.84$, $Re\approx30,000$, respectively. Fig.~\ref{fig:comp_exp_sim} compares the LES results of instantaneous vapor structures with experimental images of the corresponding operating points and shows a good agreement of the cavitation patterns. The operating point $\sigma = 1.19$ is in the regime of developing cavitation, where we observe vapor structures in the first half of the nozzle. For $\sigma = 0.84$, cavitation increases significantly and a vapor sheet extends from the nozzle inlet almost to the outlet. At $\sigma = 0.84$ the chosen inlet pressure of the simulation apparently was slightly smaller than in the experiment, resulting in a lower average velocity $\bar{u}$ ($15.2\,\si{m/s}$ instead of $15.4\,\si{m/s}$) and thus also in a slightly greater cavitation number than in the experiments. 

A validation of the simulation results can be found in \citet{Trummler:2018AAS}, where the mean streamwise velocity and the velocity fluctuations at different positions are compared with experimental data~\citep{Sou:2014hja} and show good agreement.

\subsection{Cavitation dynamics and shedding mechanisms}
In this section, we analyze the cavitation dynamics and scrutinize the shedding cycles of both operating points. Afterwards, we propose modifications to schematics of the cloud shedding process from the literature for the different shedding mechanisms. Following \citet{Stanley:2014id}, we define the beginning of a cycle by the collapse of the previously shed cloud.

\subsubsection{Developing cavitation and mainly re-entrant jet initiated shedding ($\sigma= 1.19 $) }
\label{subsec:cavdyn_sigma119}

\begin{figure}
    \centering
    \includegraphics[width=0.99\linewidth]{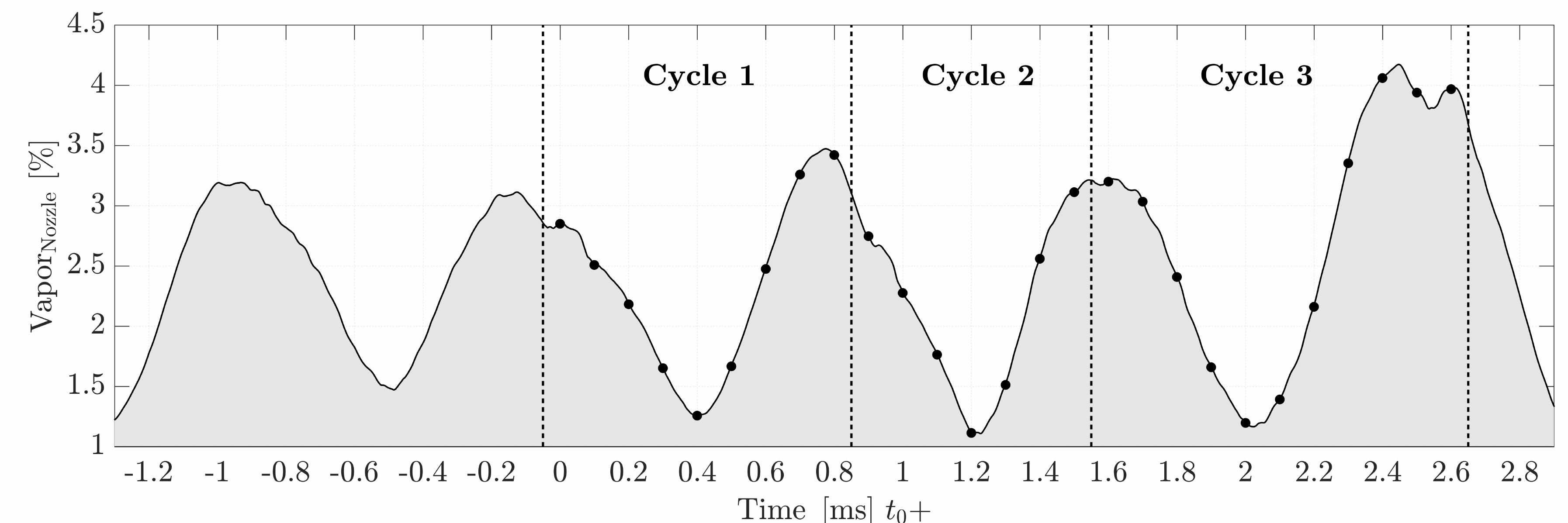}
    \caption{ 
        Vapor content over time for $\sigma= 1.19$; time steps shown in Fig.~\ref{fig:sou3_cav_dyn} are marked with a dot.
        }
 \label{fig:sou3_vap}
\end{figure}

\begin{figure}
    \centering
    \includegraphics[height=19cm]{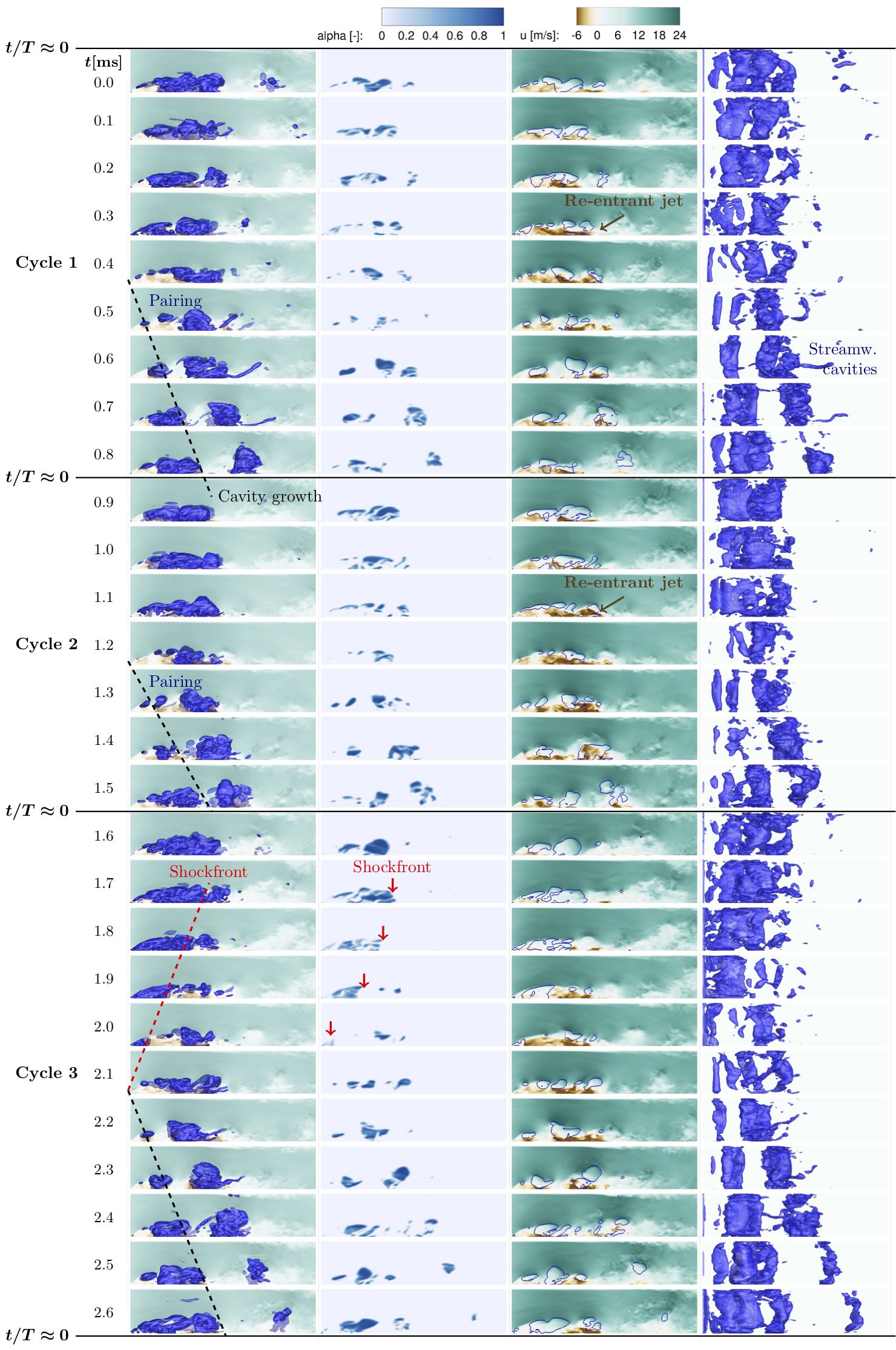}
    \caption{Time series for $\sigma= 1.19$, time step $\Delta t =0.1\; \si{ms}$, time increases from top to bottom. First to third column: side view of the midplane flow field; fourth column: top view. First column: streamwise velocity and isosurface vapor; second: vapor; third: streamwise velocity and isoline vapor; fourth: isosurface vapor. Blue isosurface or isoline 10\% vapor, black dashed line indicates the cavity growth and the red dashed one the upstream moving shock front. 
    }
    \label{fig:sou3_cav_dyn}
\end{figure}

A time series of three shedding cycles depicting instantaneous vapor structures as well as the flow field in a side and a top view is shown in Fig.~\ref{fig:sou3_cav_dyn}. The corresponding temporal evolution of the vapor content in the nozzle is shown in Fig.~\ref{fig:sou3_vap}. The dominant frequency of the integral vapor content based on spectral (Fourier) analysis of the data is $f = 1110\,\si{Hz}$ \citep{Trummler:2018AAS}, which corresponds to a period of $T = 0.9\,\si{ms}$. 

At the beginning of each shedding cycle, we observe a single cavity near the nozzle inlet, slightly shifted downstream. The shear layer at the inlet edge contains only a thin vapor sheet. In the middle of each cycle, a vapor cloud is shed and convected downstream, which manifests itself as rolling, cavitating horseshoe vortex, see e.g. fourth column of Fig.~\ref{fig:sou3_cav_dyn}. Corresponding experimental observations are given in \citet{Kubota:1989unsteady}. These detached, rolling vortices move faster than the main cavity grows, which can be clearly seen from the different trajectories in Fig.~\ref{fig:sou3_cav_dyn} (e.g. $t=2.2\,\text{--}\,2.7\,\si{ms}$). The structures become smaller as they move downstream and their velocity increases. These observations agree with experimental findings by~\citet{Kubota:1989unsteady, Sato:2002vv}. Furthermore, we observe streamwise cavitating vortices (e.g. $t = 0.6\,\si{ms}$, $t = 0.7\,\si{ms}$, $t = 2.4\,\si{ms}$). At certain time steps, these connect the main cavity with the detached structure. \citet{Saito:2003us} reported similar observations.

After the shedding, a new main cavity forms (black dashed line in Fig.~\ref{fig:sou3_cav_dyn}), and gains vapor volume by shear layer cavitation. Cavitating spanwise vortices form in the shear layer and later coalesce in a rolling motion (e.g. $t=0.4\,\text{--}\,0.5\,\si{ms}$ and $t=1.3\,\text{--}\,1.4\,\si{ms}$). Such a pairing of spanwise structures in a rolling motion was also observed in experiments~\citep{Saito:2003us, Sato:2002vv}. 

The vapor sheet is not a coherent vapor cavity, as often shown schematically~\citep{Le:1993partial,callenaere:2001,franc2004fundamentals}, but is a two-phase mixture of vapor bubbles and liquid, which also has been observed in experimental investigations~\citep{Stanley:2014id, Kubota:1989unsteady}. Furthermore, in most cases the vapor sheet is not attached to the nozzle wall but separated from the wall by a liquid film with an upstream velocity, as also reported by \citet{Stanley:2014id, Stanley:2011gr}.

The shedding during the first two cycles depicted in Fig.~\ref{fig:sou3_cav_dyn} is caused by a re-entrant jet, whereas in the last cycle it is due to a condensation shock. During the first two cycles, a thin, coherent, upstream moving liquid film at the end of the sheet can be seen over the first half of the cycle, which agrees with existing schematics of the cloud shedding process~\citep{Stanley:2014id, Le:1993partial} depicting a re-entrant jet motion over the first third of a cycle. The jet motion continues upstream and initiates a shedding in the middle of the cycle. An upstream flow underneath the cavity is present during the entire cycle since in the second half of each cycle, the rolling motion of detached vapor structures induces an upstream flow. 

During the third cycle, near wall liquid layers are not observed and the vapor reaches down to the wall, i.e. the cavity is attached, which apparently triggers the condensation shock observed during this cycle. Moving upstream the shock increases in height and eventually spans over the entire cavity height (see $t=1.9\,\si{ms}$, $t=2.0\,\si{ms}$). 

\begin{figure}
    \centering
    \includegraphics[height=19cm]{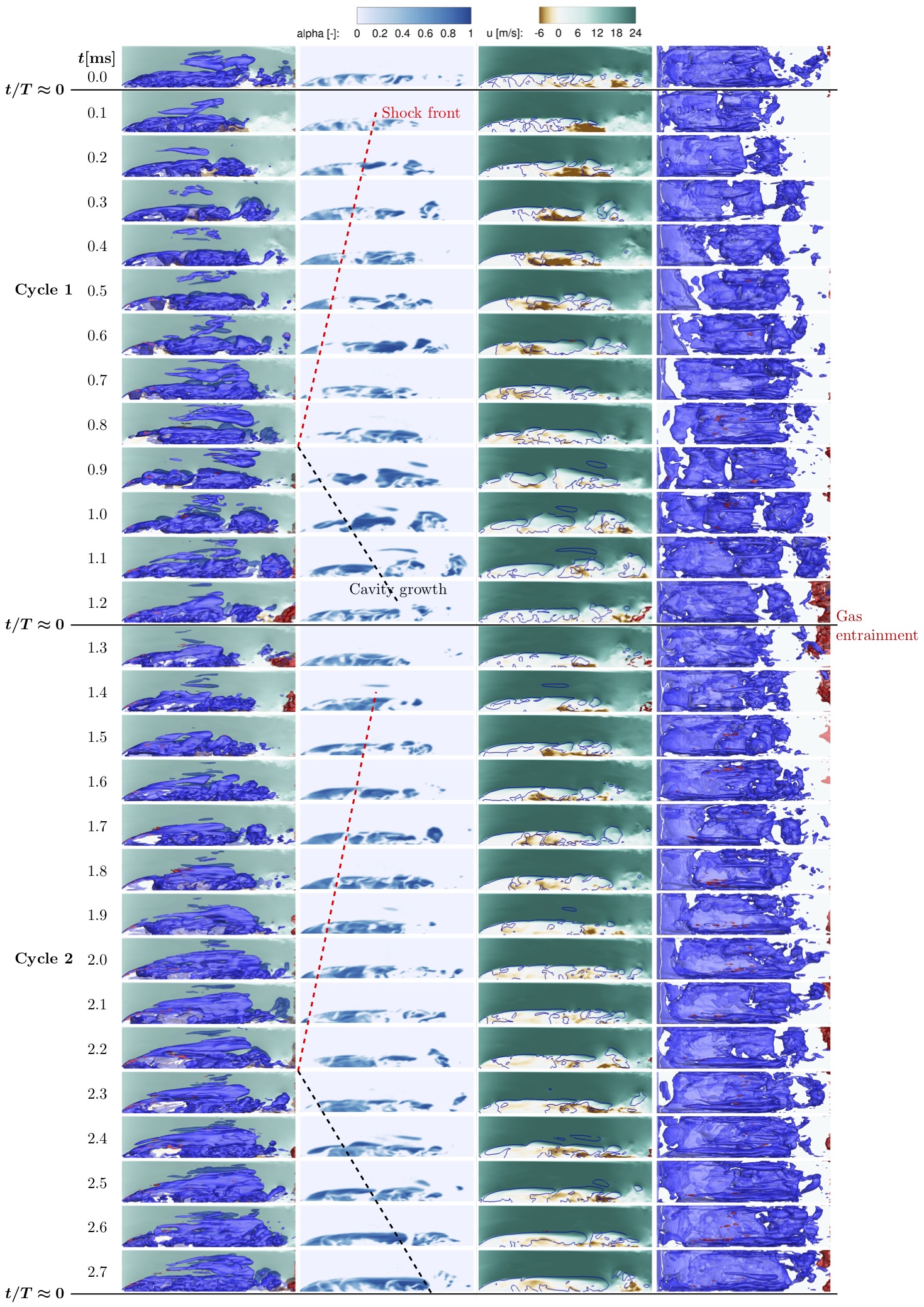}
    \caption{Time series for $\sigma= 0.84$, time step $\Delta t =0.1\; \si{ms}$, time increases from top to bottom. First to third column: side view of the midplane flow field; fourth column: top view. First column: streamwise velocity and isosurfaces vapor and gas; second: vapor; third: streamwise velocity and isoline vapor; fourth: isosurfaces vapor and gas. Blue isosurface or isoline 10\% vapor, red isosurface 10\% gas, black dashed line indicates the cavity growth and red dashed one the upstream moving shock front. 
    }
    \label{fig:sou5_cav_dyn}
\end{figure}%

\begin{figure}
    \centering
    \includegraphics[width=0.99\linewidth]{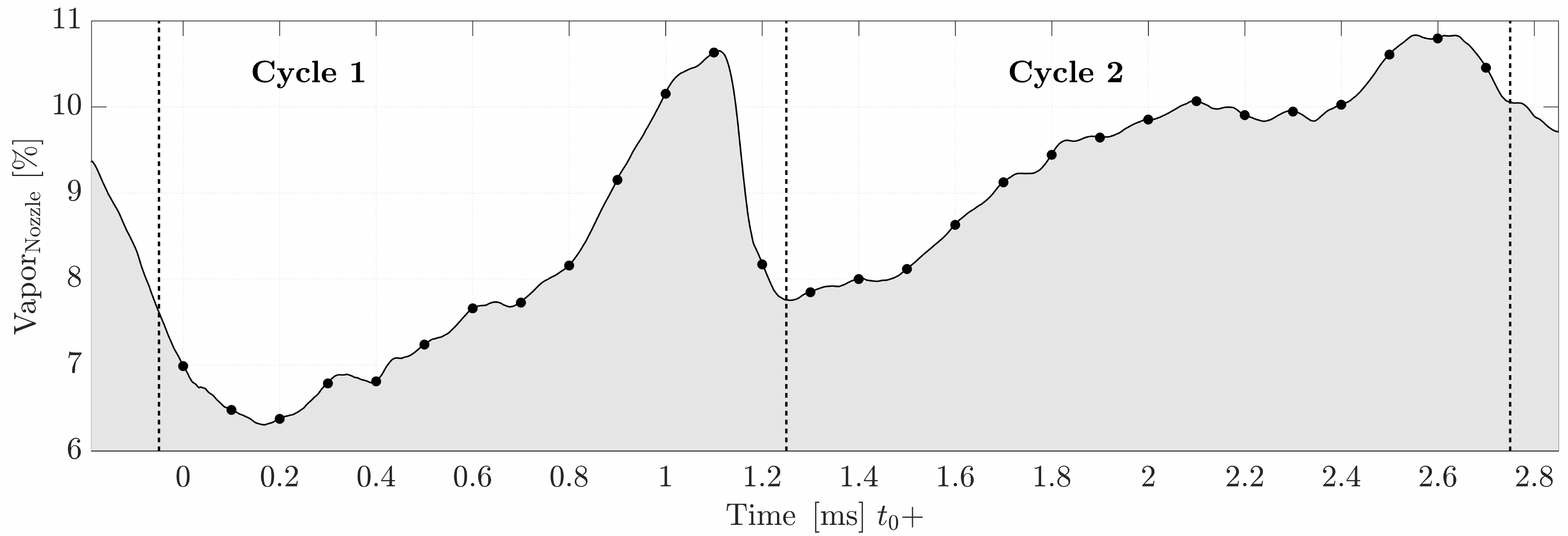}
    \caption{
         Vapor content over time for $\sigma= 0.84$; time steps shown in Fig.~\ref{fig:sou5_cav_dyn} are marked with a dot.}
    \label{fig:sou5_cav_dyn_vap}
\end{figure}%

\subsubsection{Super cavitation and condensation-shock initiated shedding ($\sigma= 0.84 $)}
\label{subsec:cavdyn_sigma084}

Fig.~\ref{fig:sou5_cav_dyn} shows a time series covering two shedding cycles. At this operating point, partial gas entrainment occurs. The red isosurface indicates a local gas content of $\geq 10\%$. Fig.~\ref{fig:sou5_cav_dyn_vap} depicts the corresponding temporal evolution of the integral vapor content. The vapor content has increased significantly and does not drop as much as for $\sigma = 1.19$. We do not observe a well-defined periodic shedding as for $\sigma = 1.19$ and the dominant frequency decreases compared to $\sigma = 1.19$, both observations also reported from experimental investigations~\citep{Stanley:2011gr}. The dominant frequency found by spectral analysis is $f = 750\,\si{Hz}$, which corresponds to $T = 1.33\,\si{ms}$. The decreasing shedding frequency with increasing cavitation is consistent with observations by experimental studies \citep{Stanley:2011gr, Ganesh:2015uj, ganesh2016bubbly}.

Streamwise vortices in the center of the nozzle can be observed at some time instants, which also have been observed in experiments~\citep{mauger2012shadowgraph} and numerical simulations~\citep{Egerer:2014wu}; the latter reference discusses the underlying mechanisms leading to the formation of these vortices. 

At the beginning of the shedding cycle, a vapor sheet spans from the inlet edge to about $3/4$ of the nozzle length. At the end of the sheet, a liquid flow with a comparably high velocity in the upstream direction is visible, similar to the observations and the schematic description by \citet{Stanley:2014id}. During the first half of the shedding cycle, smaller structures at the end of the sheet detach, are convected downstream and then collapse. Due to the detachment of these smaller structures, the cavity length is approximately constant throughout the cycle. The small detached structures are rolling, cavitating spanwise vortices. Occasionally, connecting streamwise vortices are visible, as in $t = 0.3\,\si{ms}$ or in $t = 1.7\,\si{ms}$. The re-entrant jet formed at the beginning of each cycle later transforms into a condensation shock. While the upstream moving shock leads to the condensation of the front part of the cavity, the rear part is further fed by shear layer cavitation. The complete condensation of the front part causes the detachment of the remaining rear part, see e.g. $t = 0.8\,\si{ms}$. This part is then convected further downstream and reshapes into two vapor clouds ($t=0.9\,\text{--}\,1.2\,\si{ms}$) rotating towards the nozzle outlet. The cloud further downstream reaches the outlet and leads there to gas entrainment into the nozzle ($t=1.2\,\si{ms}$, red isosurfaces). Shortly afterwards ($t=1.3\,\text{--}\,1.4\,\si{ms}$), the gas is pushed back out. However, a certain amount of gas remains in the nozzle and expands later in the low-pressure vapor region forming small gas-filled structures, see e.g. $t=1.7\,\si{ms}$ and $1.8\,\si{ms}$. The second cloud collapses close to the main cavity ($t=1.2\,\si{ms}$, $t=1.3\,\si{ms}$), and after its collapse a thin upstream liquid flow at the end of the cavity can be seen. For the second cycle, we also observe a shedding, though not visible as clearly as in the first cycle depicted, and the volume of the shed cloud is smaller. 

\subsubsection{Characterization of the shedding processes}
\label{subsubsec:govMech}

\begin{table}
\caption{
    Strouhal number ($Str$), temporal partitioning ($\lambda_{T,\phi}$) and relative velocities ($\lambda_{U,\phi}$) with $\phi=\{G,D\}$ where $G$ is growth and $D$ deformation. 'Slope' refers to the lines in the time series (Fig.~\ref{fig:sou3_cav_dyn}, Fig.~\ref{fig:sou5_cav_dyn}). 
    } 
\centering\begin{tabular}{|rcccccccccc|}
\hline
    $\boldsymbol{\sigma}$         & 
    $\boldsymbol{\bar{u}}$        &  
    $\boldsymbol{f}$                        & 
    $\boldsymbol{l_\mathrm{cav}}$  &
    $\boldsymbol{Str}$     & 
    $\boldsymbol{\lambda_\mathrm{T,G}}$ & 
    $\boldsymbol{\lambda_\mathrm{u,G}}$ & 
    $\boldsymbol{\lambda_\mathrm{u,G}^\mathrm{slope}}$ & 
    $\boldsymbol{\lambda_\mathrm{T,D}}$& 
    $\boldsymbol{\lambda_\mathrm{u,D}}$& 
    $\boldsymbol{\lambda_\mathrm{u,D-SF}^\mathrm{slope}}$ \\
    \textbf{[-]} &  \textbf{[m/s]} &   \textbf{[Hz]}  &      \textbf{[mm]} &
    \textbf{[-]} & \textbf{[-]} & \textbf{[-]} & \textbf{[-]} & 
    \textbf{[-]} & \textbf{[-]} & \textbf{[-]} \\
    \hline
    1.19  & 12.8  & 1110 & 3.6 & 0.31 & 1/2 & 0.62 & 0.53 & 1/2 & 0.62& 0.7\\
    0.84  & 15.2  &  750 & 5.8 & 0.29 & 1/3 & 0.87 & 0.86 & 2/3 & 0.44& 0.33\\
    \hline
    \end{tabular}%
  \label{tab:govmech}%
\end{table}%

The shedding at $\sigma=1.19$ is mainly caused by re-entrant jet motion whereas at $\sigma=0.84$ condensation shocks are predominant. The dominance of condensation shocks for the lower cavitation number is in accordance with experimental investigations~\citep{Jahangir:2018id, Arndt:2001tx}. E.g.~\citet{Jahangir:2018id} determined that for a convergent, divergent nozzle the shedding is shock dominated for $\sigma < 0.75$. The comparison to our results indicates that for our nozzle geometry the limits are shifted to higher cavitation numbers, which may result from a stronger blocking effect due to the constant cross-section. \citet{Ganesh:2015uj} found that the formation of condensation shocks is favored at increased void fractions. 

The dimensionless Strouhal number for shedding processes is
    \begin{equation}
        Str=f\cdot l_\mathrm{cav}/\bar{u},  
        \label{eq:Str}
    \end{equation}
where $l_\mathrm{cav}$ the length of the main cavity, here evaluated in the time series. The values obtained are $Str \approx 0.3$ (see Table~\ref{tab:govmech}) and compare well with experimental values for cylindrical orifices with $Str = 0.3\,\text{--}\,0.5$~\citep{Stanley:2011gr, Sato:2002vv, Sugimoto:2009}.

As can be seen in the time series (Fig.~\ref{fig:sou3_cav_dyn}, Fig.~\ref{fig:sou5_cav_dyn}), shedding cycles can be divided into the upstream motion of a disturbance ($\phi = D$), which is either a re-entrant jet ($RJ$) or a condensation shock ($CS$), and the growth phase of the main cavity ($\phi = G$) with the temporal partitioning of
    \begin{equation}
        \lambda_{T,\phi}=t_{\phi}/T\, \;\;\;\; \mathrm{and}\;\;\;\; \sum_{\phi}{\lambda_{T,\phi}}=1
        \label{eq:lambda_t}
    \end{equation}
where $t_{\phi}$ stands for the time the process $\phi=\{D,G\}$ takes. We observe for the re-entrant jet governed cycles at $\sigma=1.19$ a duration of about half a period for each process and thus $\lambda_{T,G}=\lambda_{T,D-RJ}=1/2$ (see Fig.~\ref{fig:sou3_cav_dyn}). Our $\lambda_{T,D-RJ}$ is slightly higher than reported in the literature as \citet{Le:1993partial} proposed $\lambda_{T,D-RJ}=1/3$ and \citet{callenaere:2001} found $\lambda_{T,D-RJ}\approx0.40$ and a length of $\lambda_{L,D-RJ}=0.75$. For the shock governed regime, the motion of the disturbance is slower and takes about $\lambda_{T,D-SF}=2/3$ (see Fig.~\ref{fig:sou5_cav_dyn}). A dependency of the temporal partitioning on the governing shedding mechanism can also be seen in the time series in Fig.~\ref{fig:sou3_cav_dyn}, where the last cycle, with the condensation shock, is significantly longer than the previous ones. In the data provided by \citet{budich2018jfm} for a condensation shock governed shedding of a cavitating flow over a wedge, the velocity ratios are $|u_{G}/u_{D-SF}|= 1.22$ which implies that $\lambda_{T,G}<\lambda_{T,D-SF}$ and confirms our observation. 

The velocity ($u_{\phi}=\lambda_{U,\phi}\,\bar{u}$) of each process is the ratio of the length $l_{\phi}$, which can be expressed as $l_{\phi}=\lambda_{L,\phi}\,l_{cav}$, to the time $t_{\phi}$ as 
\begin{equation}
        u_{\phi}=\frac{l_{\phi}}{t_{\phi}}=\frac{\lambda_{L,\phi}\,l_{cav}}{\lambda_{T,{\phi}}\,T}, 
        \label{eq:u_est1}
    \end{equation}
with $\lambda_{L,\phi}\approx 1$ and Eq.~\eqref{eq:Str} we obtain for the dimensionless velocity 
    \begin{equation}
        \lambda_{u,\phi}\approx \frac{Str}{\lambda_{T,\phi}}. 
        \label{eq:u_est2}
    \end{equation}
Table~\ref{tab:govmech} compares the velocities estimated using Eq.~\eqref{eq:u_est2}, which represent time averaged values, and the ones obtained by the mean slope of the lines tracing the cavity end and the shock front in the time series (Fig.~\ref{fig:sou3_cav_dyn}, Fig.~\ref{fig:sou5_cav_dyn}). The agreement of these values shows that Eq.~\eqref{eq:u_est2} provides a good estimate. Furthermore, the velocities determined and estimated for the disturbance $\lambda_{U,D}$, $\lambda_{U,D}^\mathrm{slope}$ are in accordance with the data from the literature. For re-entrant jets, values of approximately 0.5 $\bar{u}$ (e.g.~\citet{Pham:1999tp,callenaere:2001}) are given, which is within the range of our results. Detailed analyses follow in subsection~\ref{subsec:flowfield}. For the shock front velocities values of 35~\text{--}60\% of the free flow velocity are reported by \citet{Ganesh:2015uj} for cavitating flows over a wedge. For a cavitating nozzle flow, \citet{Stanley:2014id} observed an upstream moving deformation at the interface of the lower cavity at a velocity of 30~\text{--}~80\% $\bar{u}$ for the range of our cavitation numbers. Both ranges of the shock front velocity agree to our results of $\lambda_{U,D-SF}^\mathrm{slope}=0.33\,\text{--}\,0.7$, see Table~\ref{tab:govmech}. 

\begin{figure}
    \centering
    \subfigure[]{\includegraphics[width=0.44\linewidth]{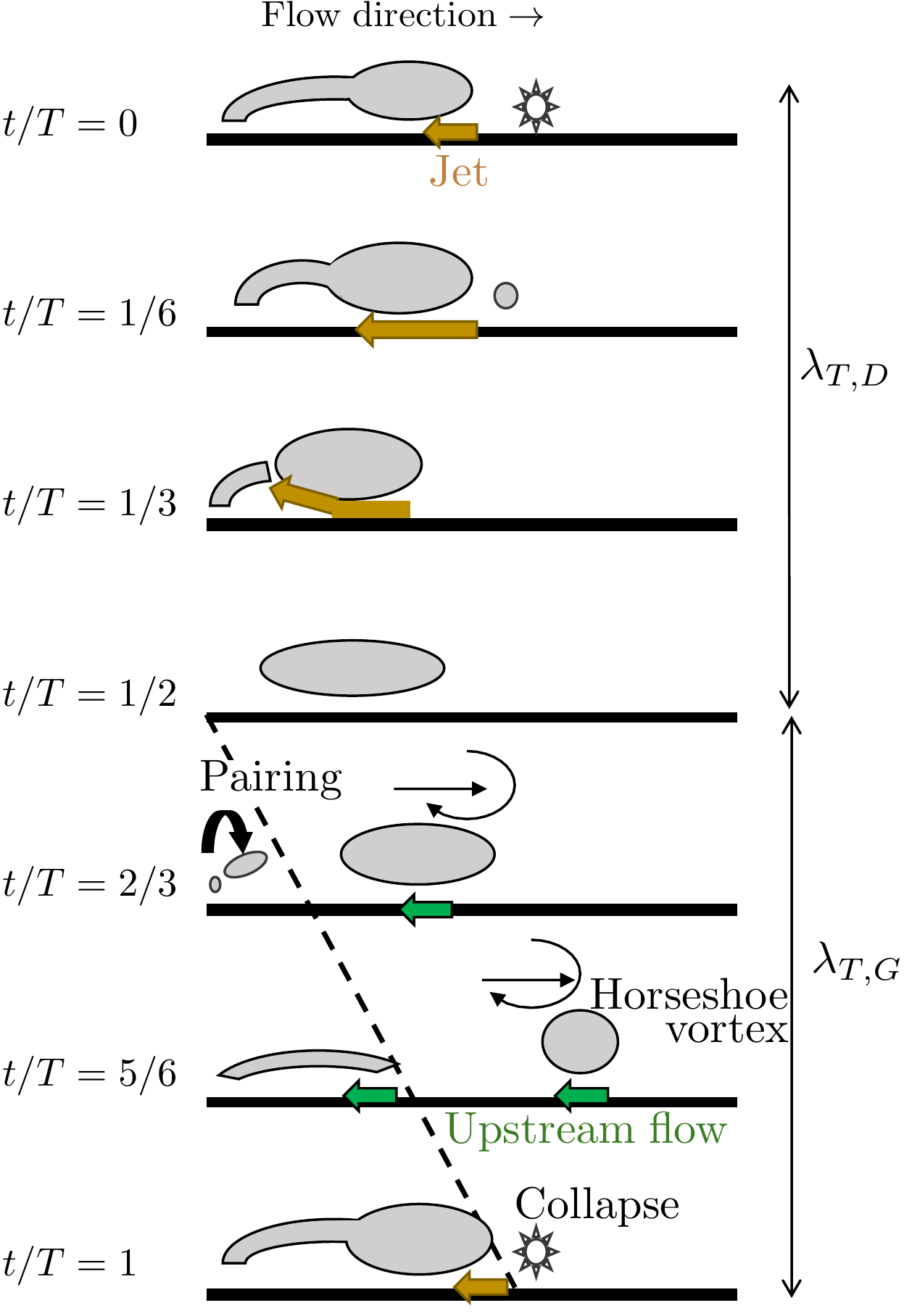}}
    \hspace{1cm}
    \subfigure[]{\includegraphics[width=0.44\linewidth]{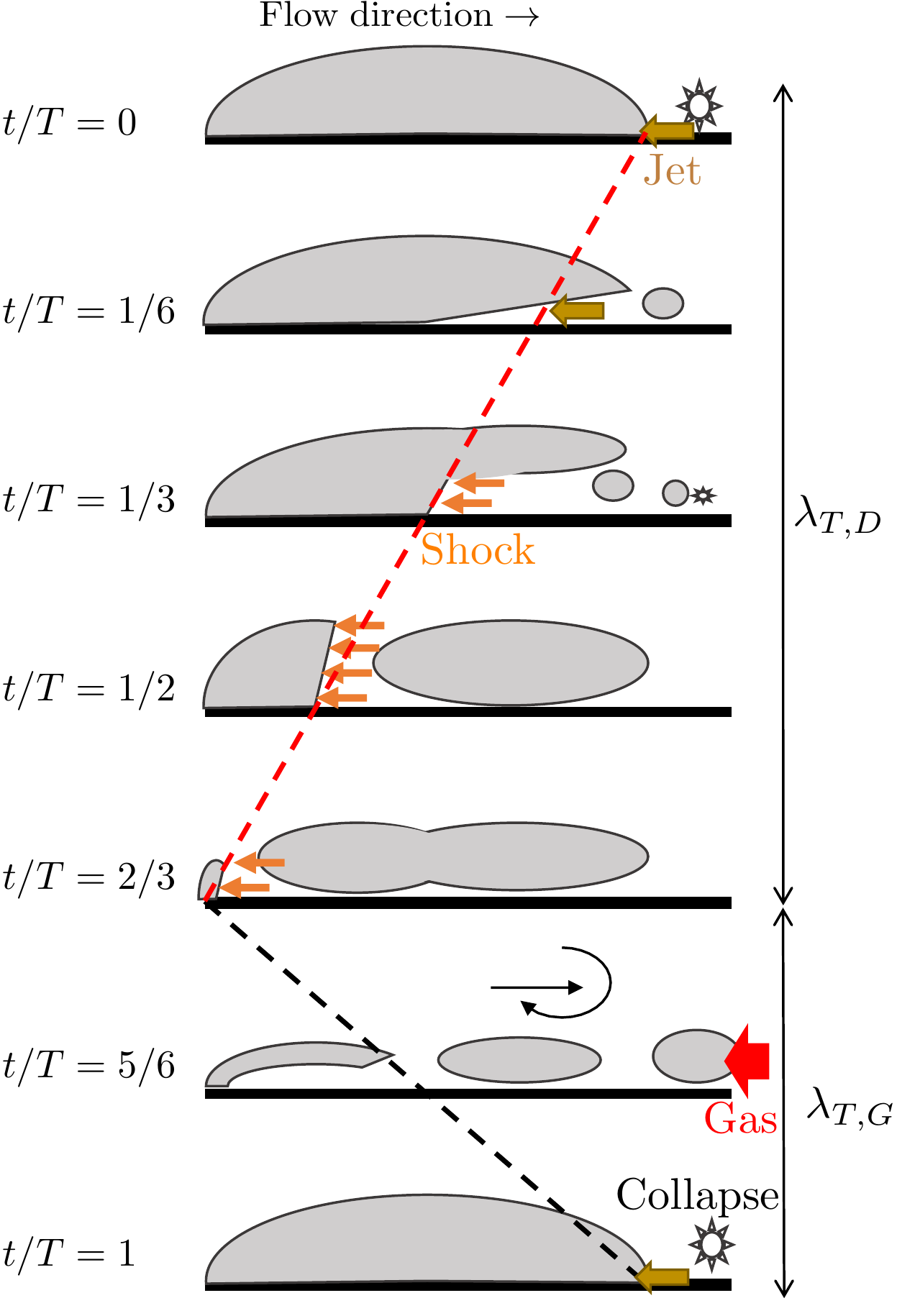}}
    \caption{
    Schematics of cloud shedding processes adapted and modified from those proposed by \citet{Stanley:2014id, Le:1993partial} for re-entrant jet governed cloud shedding. (a) developing cavitation with re-entrant jet initiated shedding (b) super cavitation with condensation shock initiated shedding. Flow direction is from left to right; vapor appears gray. }
    \label{fig:scheme2}
\end{figure}%

\subsubsection{Schematics of the cloud shedding process}
Based on the previous observations we have adapted existing schematics of the cloud shedding process~\citep{Stanley:2014id, Le:1993partial} as shown in Fig.~\ref{fig:scheme2} and propose modifications depending on the governing shedding mechanism. 

For the re-entrant jet initiated shedding at the developing cavitation, an upstream motion of a liquid jet is present during the first half of a cycle. In the middle of the shedding cycle, a vapor cloud is shed and convected downstream as a rolling horse-shoe vortex. The growth of the new main cavity is due to shear layer cavitation and paring and coalescence of the formed spanwise vortices. An upstream, liquid, near-wall flow is present throughout the cycle - first due to the re-entrant jet motion and later due to the rolling motion of the detached structures. 

In case of a shock initiated shedding the vapor sheet is attached to the nozzle wall. At the beginning of a cycle, a re-entrant jet forms at the end of the sheet and moves upstream transforming into a condensation shock. Here the temporal partitioning into the convection of the disturbance is longer than for the re-entrant jet initiated shedding, as also discussed above. The condensation of the upstream part leads to the detachment of the rear part. In this cavitation regime, detached vapor clouds reaching the nozzle outlet can lead to gas entrainment. In this study, the gas entrainment affects only the flow field directly at the nozzle outlet, and we did not observe a direct impact on cavitation dynamics and shedding.

\subsection{Flow field}
\label{subsec:flowfield}

In this subsection special emphasis is placed on the near-wall upstream flow and the investigation of the underlying mechanisms leading to the re-entrant jet formation. 

\subsubsection{Instantaneous flow field and vorticity}
\label{subsubsec:vort}

\begin{figure}
    \centering
    \subfigure{\includegraphics[width=\linewidth]{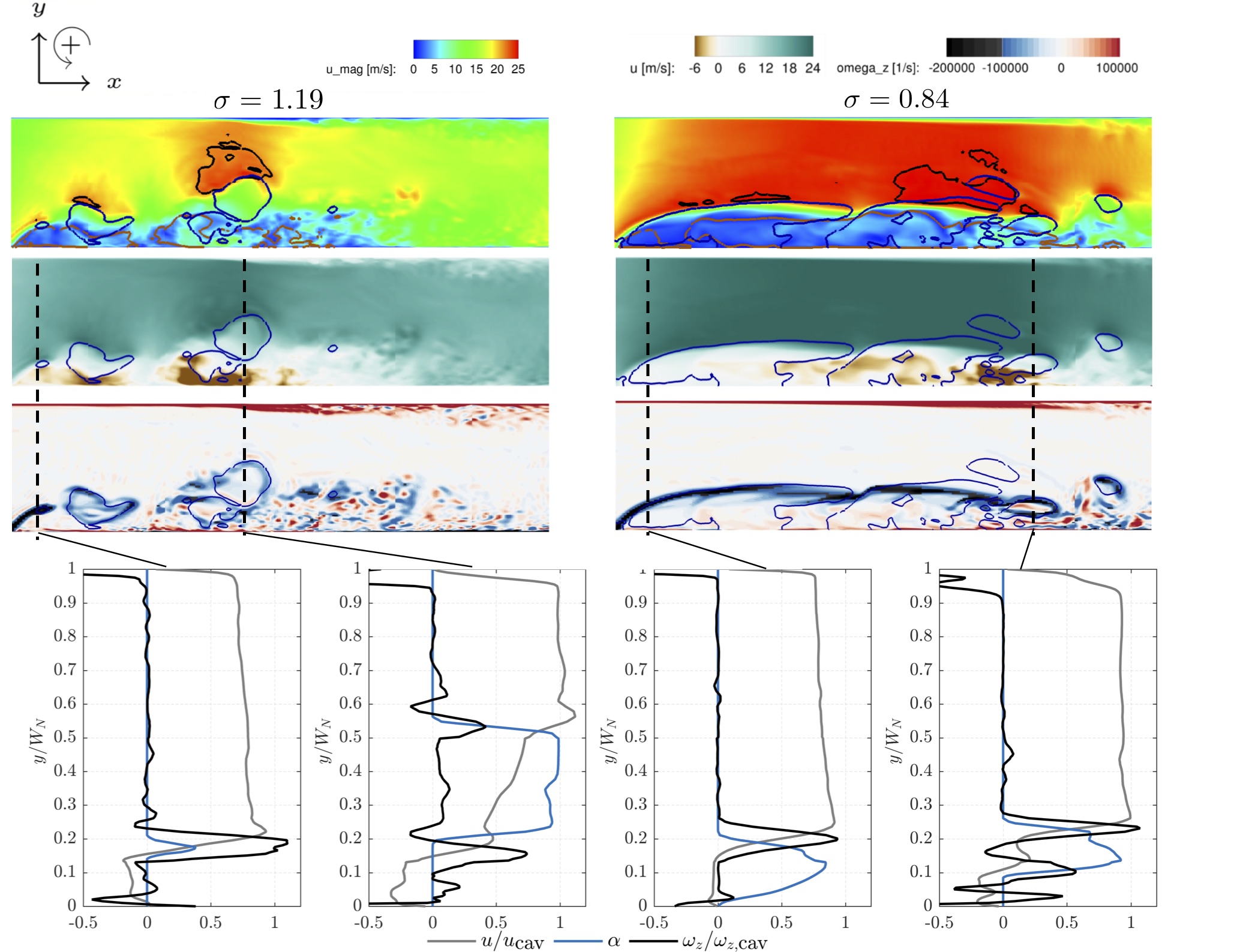}}
    \caption{
    Flow field and vorticity on the midplane ($x\,-\,y$). Left for $\sigma=1.19$ at $t=2.3\,\si{ms}$, right for $\sigma=0.84$ at $t=2.5\,\si{ms}$. First row: velocity magnitude and isolines $u_\mathrm{cav}$ (black), $u=0\,\si{m/s}$ (brown) and $\alpha=10\%$ (blue); second: streamwise velocity and isoline $\alpha=10\%$ (blue); third: vorticity and isoline $\alpha=10\%$ (blue); at the bottom: extracted data at the marked positions. }
\label{fig:vort2}
\end{figure}%

The instantaneous flow field and the vorticity distribution on the midplane for one time step of each operating point are shown in Fig.~\ref{fig:vort2}. In addition, data extracted at the nozzle inlet and across a detached vapor structure are plotted at the bottom. In Fig.~\ref{fig:vort2}, the flow detaches at the nozzle inlet and the highest velocity occurs in the detached shear layer, where the liquid starts to cavitate. The velocity above the cavity ($u_\mathrm{cav}$) can be approximated by Bernoulli's equation applied to a streamline from the inlet to the cavity 
    \begin{equation}
        u_\mathrm{cav}=\sqrt{2\,\cdot\,(p_\mathrm{total,inlet}-p_\mathrm{sat})/\rho}.
        \label{eq:u_max}
    \end{equation}
$u_\mathrm{cav}$ is $21.66\,\si{m/s}$ and $24.54\,\si{m/s}$ for $\sigma=1.19$ and $\sigma=0.84$ respectively, which closely matches the velocities at the cavity interface in Fig.~\ref{fig:vort2}. At $\sigma=1.19$ the velocity on the upper side of the vapor structures increases locally due to displacement, whereas at $\sigma = 0.84$ the velocity above the detached shear layer is approximately constant. 

The flow around the cavity induces a negative circulation $\Gamma_\mathrm{cav}$ in $z$ direction. Circulation $\Gamma$ is the line integral of the velocity $\underline{u}$ along a closed line $\underline{s}$ 
\begin{equation}
    \Gamma=\oint_{\partial s} \,\underline{u}\,d\underline{s}
    \label{eq:gamma1}
\end{equation}
or, due to Stokes' theorem, the integral of the vorticity $\underline{\omega}=\nabla \times \underline{u}$ over the enclosed area $A$ with the normal vector $\underline{n}$ as 
\begin{equation}
    \Gamma=\int_{A} (\nabla \times \underline{u})\cdot \underline{n} \, d A = \int_{A} \underline{\omega}\cdot \underline{n} \, d A .  
    \label{eq:gamma2}
\end{equation}
$\Gamma_\mathrm{cav}$ can be estimated by using Eq.~\eqref{eq:gamma1}
with an integration path along the cavity interface, see e.g.~\citet{Le:1993partial}, as  
\begin{equation}
    |\Gamma_\mathrm{cav}| \approx u_\mathrm{cav}\cdot l_\mathrm{cav}.   
    \label{eq:gamma3}
\end{equation}
Fig.~\ref{fig:vort2} shows that at $\sigma=0.84$ the vorticity $\omega_{z}$ for this circulation is concentrated in a thin layer at the boundary of the cavity. Thus, $\Gamma_\mathrm{cav}$ can be approximated with Eq.~\eqref{eq:gamma2} by the vorticity in this layer $\omega_{z, \mathrm{cav}}$ and the area, where it is present ($l_\mathrm{cav} \cdot h_\mathrm{vort}$ ), as 
\begin{equation}
    \Gamma_\mathrm{cav}\approx \omega_{z,\mathrm{cav}} \cdot l_\mathrm{cav}\cdot h_\mathrm{vort}
    \label{eq:gamma4}
\end{equation}
and with Eq.~\eqref{eq:gamma3} we obtain
\begin{equation}
    \omega_{z,\mathrm{cav}} \approx - u_\mathrm{cav}/h_\mathrm{vort}. 
    \label{eq:gamma5}
\end{equation}
We estimate $h_\mathrm{vort}\approx 0.1\,\si{mm}$ and the derived values for $\omega_{z,\mathrm{cav}}$ correspond at both operating points to the vorticity at the shear layer at the nozzle inlet, see Fig.~\ref{fig:vort2}. 

At $\sigma=1.19$, the highly vortical shear layer transitions into spanwise vortices, which are surrounded by vorticity of reduced magnitude. At both operating points, the shed vapor structures downstream are framed by layers of negative vorticity of about $0.5-1 \,\omega_{z,\mathrm{cav}}$. Within these structures, the magnitude of the vorticity is either significantly decreased ($\sigma=1.19$) or decreases radially ($\sigma = 0.84$), indicating that vapor structures do not rotate like a rigid body. Note that at $\sigma=1.19$ the data is extracted across a vapor structure which is pairing in a rolling motion with another one. 

Underneath the detached vapor structures at $\sigma=1.19$ and at the end of the attached sheet of $\sigma = 0.84$, there is a strongly pronounced upstream flow of about $0.3\,u_\mathrm{cav}$ that correlates to a high negative vorticity at these locations.

Additionally, we want to emphasize here the strong interaction of vortical structures and cavitation. Cavitation often occurs in the low pressure cores of these structures. The time series in Fig.~\ref{fig:sou3_cav_dyn} and \ref{fig:sou5_cav_dyn} show streamwise and spanwise vortical structures. Vortex cavitation is assessed in detail in e.g.~\citet{arndt2002cavitation, ji2015large, ji2014numerical, cheng2020large}. 

\subsubsection{Near-wall upstream flow}
\label{subsec:jet} 

\begin{figure}
  \centering
  \includegraphics[width=\linewidth]{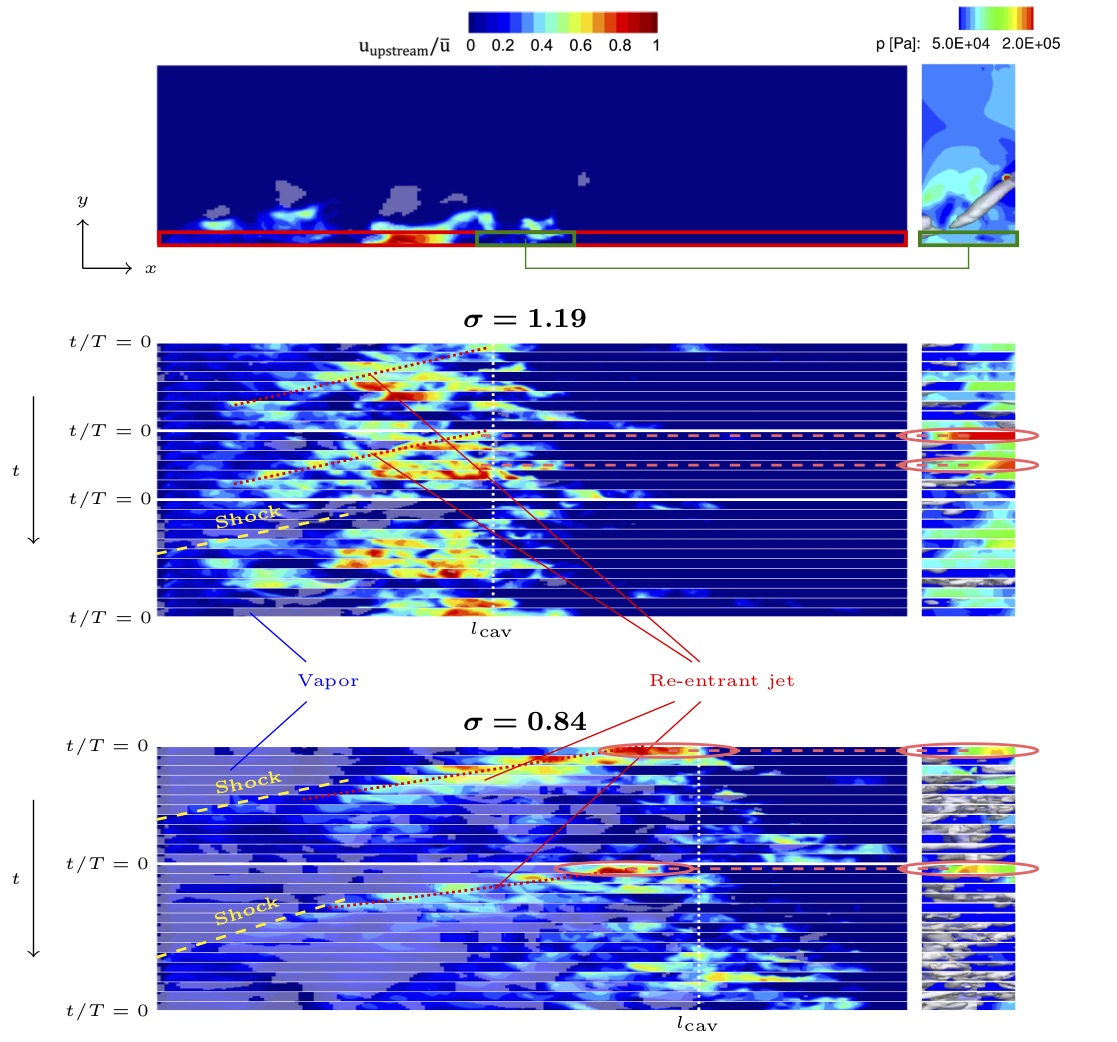}
  \caption{Time series ($x$-$t$) of streamwise velocity $u$ (left) and pressure field $p$ (right) close to the wall ($h=0.1\,\si{mm}$). Vapor regions ($\alpha=0.1$) are indicated as bright regions (left) and as white isosurface(right). The top panel illustrates the investigated regions. Red dotted lines indicate the re-entrant jet motion and yellow dashed lines highlight the shock front. Pressure peaks downstream of the cavity are correlated to the velocity field and are marked in orange. }
  \label{fig:x_t_diagram}
\end{figure}%

\begin{figure}
  \centering
  \subfigure{\includegraphics[width=0.8\linewidth]{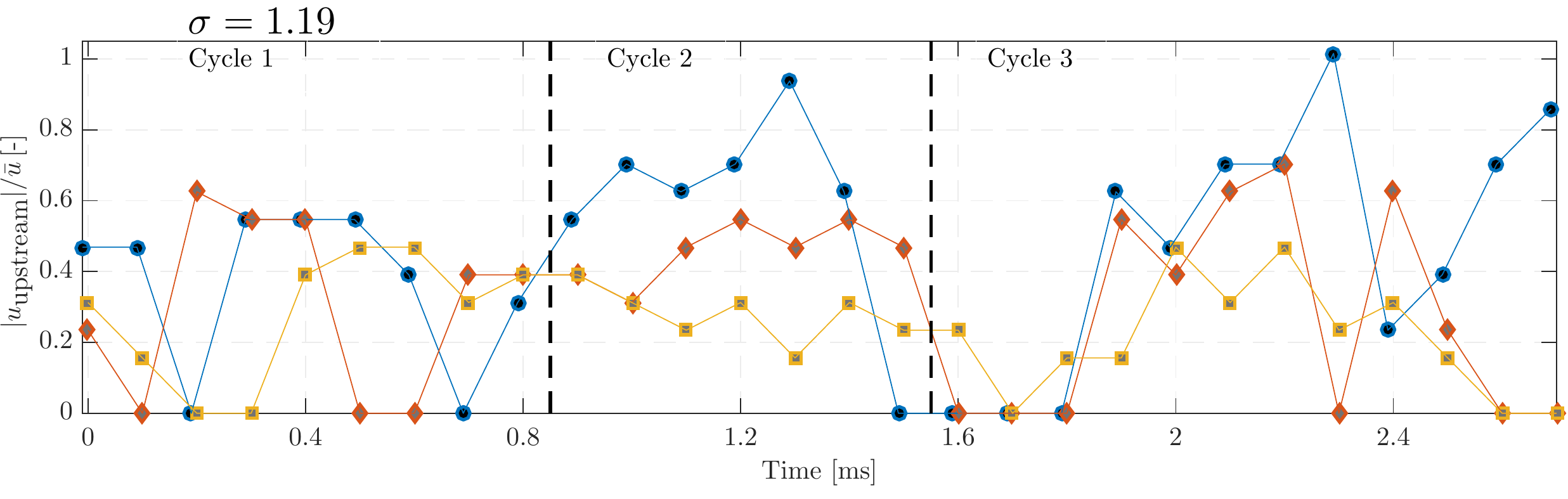}}
  \subfigure{\includegraphics[width=0.8\linewidth]{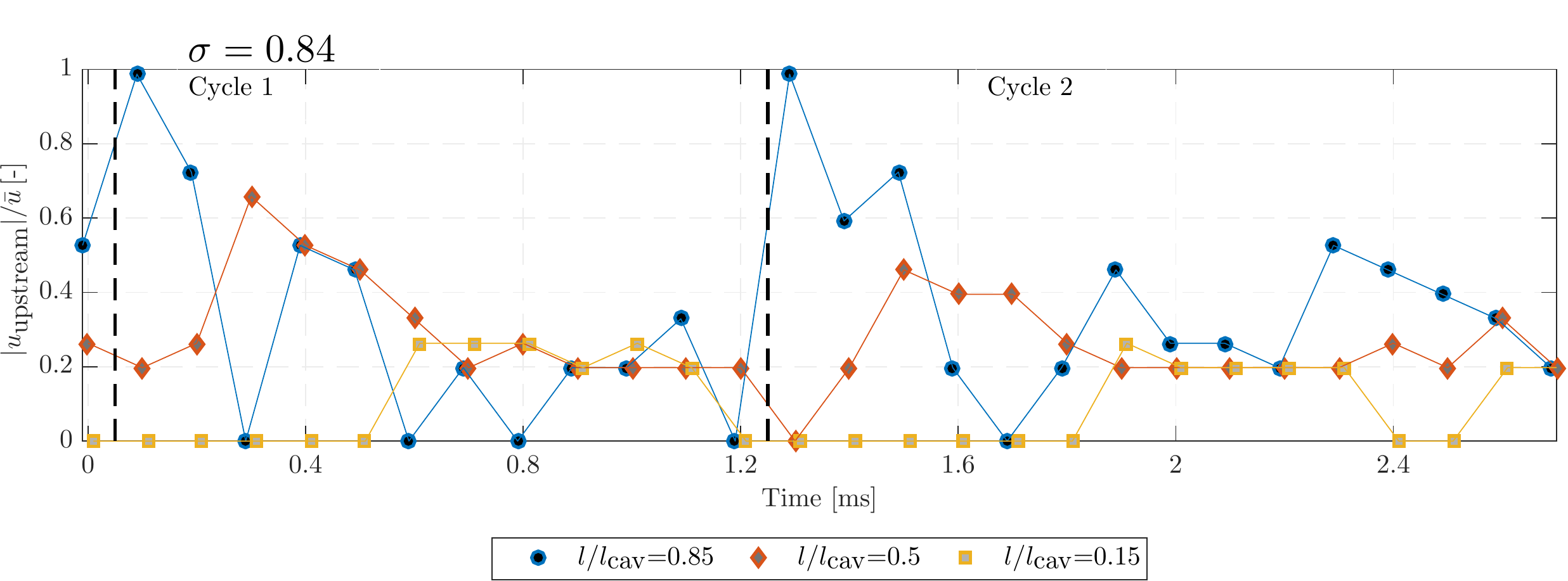}}
  \caption{Near-wall upstream velocity normalized by mean nozzle velocity $|u_\mathrm{upstream}|/\bar{u}$ evaluated at different positions as a function of time.}
  \label{fig:re_jet_vel}
\end{figure}%

We analyze the near-wall upstream flow using $x$-$t$ diagrams in which the flow field on the midplane close to the wall is extracted and arranged with increasing time, see Fig.~\ref{fig:x_t_diagram}. Furthermore, the pressure field at the end of the cavitation zone is depicted in the second column in Fig.~\ref{fig:x_t_diagram}. The upstream velocity at $l/l_{cav}=\{0.15, 0.5, 0.85\}$ is plotted over time in Fig.~\ref{fig:re_jet_vel}.

In the $x$-$t$ diagram of $\sigma = 1.19$ the formation of a re-entrant jet at the beginning of each cycle and its upstream movement can be seen. The highest velocity peaks occur at the end of the sheet at about $2/3$ of each cycle, see also Fig.~\ref{fig:re_jet_vel}, and are apparently induced by the rolling motion of the detached structures, see Fig.~\ref{fig:sou3_cav_dyn} and subsection~\ref{subsubsec:vort}. Additionally, this upstream flow can be amplified by the pressure peak induced by the collapse of smaller vapor structures at the end of the sheet, see middle of second cycle. Furthermore, the extracted velocities (Fig.~\ref{fig:re_jet_vel}) illustrate that the velocity magnitude tends to decrease in upstream direction.

For $\sigma = 0.84$, the $x$-$t$ diagram (Fig.~\ref{fig:x_t_diagram}) and the extracted velocity (Fig.~\ref{fig:re_jet_vel}) show a clear and unique peak at the end of the cavitation zone at the beginning of the two cycles, which corresponds to the re-entrant jet formed there. A comparison with the pressure field (left column of Fig.~\ref{fig:x_t_diagram}) reveals that negative velocity peaks correlate with pressure peaks induced by the collapse of the shed cloud. The fluid with peak velocity moves upstream and slows down. 
In the $x$-$t$ diagram it can be seen that further upstream the jet and the upstream flow are not purely liquid anymore and the jet also extends into the vapor region. 
While moving upstream the jet transforms into a condensation shock, which is clearly visible by the upstream moving vapor front in the $x$-$t$ diagram. Close to the nozzle inlet (see 0.15~$l_\mathrm{cav}$ in Fig.~\ref{fig:re_jet_vel}), there is only an upstream flow after the condensation shock has passed. In the second half of the cycles, the velocity magnitudes at all three positions are about the same (20 - 30\% of $\bar{u}$), which agrees with the measured data of \citet{Stanley:2014id}. We point out that \citet{Stanley:2014id} reported an upstream moving deformation, which we believe is related to a condensation shock. 

The re-entrant jet corresponds, in our notation of the shedding cycles, to the upstream flow at the beginning of the cycle. The determined re-entrant jet velocity magnitudes have peaks in the range of the mean velocity $\bar{u}$ at the end of the cavity and decrease in magnitude in upstream direction, which is in good agreement with experimental observations~\citep{Pham:1999tp,Sakoda:2001tn}. At $\sigma = 1.19$, the re-entrant jet velocities are slightly lower and match well with the re-entrant jet velocity of half the mean velocity determined by~\citet{callenaere:2001}. 

So far, the driving mechanism behind the re-entrant jet has not yet been completely clarified. In our simulations we observe for $\sigma = 0.84$ a clear correlation between the collapse of the shed cloud and a more pronounced upstream flow, which was proposed in the literature by e.g.~\citet{Leroux:2004jl,Leroux:2005ii}. However, at $\sigma = 1.19$ we observe this correlation only once, which matches the finding by \citet{CoutierDelgosha:2007dp} that this relation is flow condition dependent. Moreover, we found that at $\sigma = 1.19$ the highest magnitude of the upstream flow is induced by the rolling motion of detached vapor structures and not associated to the re-entrant jet motion. 

\section{Conclusion}
\label{sec:Conclusion}

Cloud cavitation is a common yet challenging phenomenon that occurs in external and in internal cavitating flows. Although orifices and nozzles with constant cross-section play an important role in technical applications, detailed investigations of cavitation dynamics and shedding mechanisms have been far less performed than for external flows or for flows in convergent-divergent geometries. 

In this paper, cavitation dynamics and shedding mechanisms of cavitating nozzle flows were investigated using wall-resolved LES results. The three-dimensional flow field data of several shedding cycles provide a deep insight into the physical processes of the cloud cavitation mechanisms in nozzle flows and allow detailed analyses of the flow field. Our results compare well with the reference experiment and reported observations for cavitation dynamics. 

In our simulations, the shedding for inertia-driven super cavitation ($\sigma = 0.84$) is initiated by condensation shocks, whereas for developing cavitation ($\sigma = 1.19$) it is primarily initiated by re-entrant jets. The occurrence of condensation shocks at lower cavitation numbers is consistent with experimental findings. Our investigation reveals for the first time condensation shocks in cavitating nozzle flows with constant cross-section. Based on our observations, we have analyzed the shedding in detail and then adapted and extended the schematics of the cloud shedding process from the literature for the different mechanisms. 

At both operating points, a re-entrant jet forms at the beginning of the cycle and in case of shock-initiated shedding the jet transforms further upstream into a shock. Our results confirm the existing theory that the re-entrant jet formation is related to the pressure peak induced by the collapse of the detached cloud. The determined re-entrant jet velocities are in the order of the mean nozzle velocity and decrease in upstream direction. 
For developing cavitation a pronounced upstream flow is present during the entire cycle, which first can be characterized as re-entrant jet motion and later as an upstream flow induced by the rolling motion of the detached vapor structures, which lead to the highest upstream velocities. 

The analysis performed contributes to an improved understanding of cavitation dynamics in nozzles and has demonstrated the occurrence of condensation shocks there. Furthermore, this paper provides detailed data on the shedding processes and the near-wall upstream flow and therewith complements the existing literature with data for cavitating flows in a rectangular step nozzle with injection into gas. 

\section*{Acknowledgments}
The authors acknowledge the Gauss for Supercomputing e.V. for granting computational resources on the GCS Supercomputer SuperMUC at the Leibniz Supercomputing Centre (LRZ, www.lrz.de). Nikolaus A. Adams acknowledges support through the ERC Advanced Grant NANOSHOCK (2015).

\bibliography{main}

\begin{thebibliography}{70}
\expandafter\ifx\csname natexlab\endcsname\relax\def\natexlab#1{#1}\fi
\expandafter\ifx\csname url\endcsname\relax
  \def\url#1{\texttt{#1}}\fi
\expandafter\ifx\csname urlprefix\endcsname\relax\def\urlprefix{URL }\fi

\bibitem[{Adams et~al.(2004)Adams, Hickel, and Franz}]{adams2004implicit}
Adams, N., Hickel, S., Franz, S., 2004. Implicit subgrid-scale modeling by
  adaptive deconvolution. Journal of Computational Physics 200~(2), 412--431.

\bibitem[{Arndt(2002)}]{arndt2002cavitation}
Arndt, R.~E., 2002. Cavitation in vortical flows. Annual review of fluid
  mechanics 34~(1), 143--175.

\bibitem[{Arndt et~al.(2001)Arndt, Song, Kjeldsen, and Keller}]{Arndt:2001tx}
Arndt, R. E.~A., Song, C. C.~S., Kjeldsen, M., Keller, A., 2001. Instability of
  partial cavitation: A numerical/experimental approach. In: Twenty-Third
  Symposium on Naval Hydrodynamics Office of Naval Research Bassin d'Essais des
  Carenes National Research Council. pp. 599--615.

\bibitem[{Asi(2006)}]{asi2006failure}
Asi, O., 2006. Failure of a diesel engine injector nozzle by cavitation damage.
  Engineering Failure Analysis 13~(7), 1126--1133.

\bibitem[{Beattie and Whalley(1982)}]{Beattie:1982ut}
Beattie, D. R.~H., Whalley, P.~B., 1982. A simple two-phase frictional pressure
  drop calculation method. International Journal of Multiphase Flow 8~(1),
  83--87.

\bibitem[{Beban et~al.(2017)Beban, Schmidt, and Adams}]{beban2017numerical}
Beban, B., Schmidt, S., Adams, N., 2017. Numerical study of submerged
  cavitating throttle flows. Atomization and Sprays 27~(8), 723--739.

\bibitem[{Bi{\c c}er and Sou(2015)}]{Bicer:2015cp}
Bi{\c c}er, B., Sou, A., 2015. Numerical models for simulation of cavitation in
  diesel injector nozzles. Atomization and Sprays 25~(12), 1063--1080.

\bibitem[{Bi{\c{c}}er and Sou(2016)}]{Bicer:2017ia}
Bi{\c{c}}er, B., Sou, A., 2016. Application of the improved cavitation model to
  turbulent cavitating flow in fuel injector nozzle. Applied Mathematical
  Modelling 40~(7-8), 4712--4726.

\bibitem[{Budich et~al.(2018)Budich, Schmidt, and Adams}]{budich2018jfm}
Budich, B., Schmidt, S.~J., Adams, N.~A., 2018. Numerical simulation and
  analysis of condensation shocks in cavitating flow. Journal of Fluid
  Mechanics 838, 759--813.

\bibitem[{Callenaere et~al.(2001)Callenaere, Franc, Michel, and
  Riondet}]{callenaere:2001}
Callenaere, M., Franc, J.-P., Michel, J.-M., Riondet, M., 2001. The cavitation
  instability induced by the development of a re-entrant jet. Journal of Fluid
  Mechanics 444, 223--256.

\bibitem[{Chaves et~al.(1995)Chaves, Knapp, Kubitzek, Obermeier, and
  Schneider}]{Chaves:1995vt}
Chaves, H., Knapp, M., Kubitzek, A., Obermeier, F., Schneider, T., 1995.
  Experimental study of cavitation in the nozzle hole of diesel injectors using
  transparent nozzles. SAE International, 645--657.

\bibitem[{Cheng et~al.(2020)Cheng, Bai, Long, Ji, Peng, and
  Farhat}]{cheng2020large}
Cheng, H., Bai, X., Long, X., Ji, B., Peng, X., Farhat, M., 2020. Large eddy
  simulation of the tip-leakage cavitating flow with an insight on how
  cavitation influences vorticity and turbulence. Applied Mathematical
  Modelling 77, 788--809.

\bibitem[{Coutier-Delgosha et~al.(2007)Coutier-Delgosha, Stutz, Vabre, and
  Legoupil}]{CoutierDelgosha:2007dp}
Coutier-Delgosha, O., Stutz, B., Vabre, A., Legoupil, S., 2007. Analysis of
  cavitating flow structure by experimental and numerical investigations.
  Journal of Fluid Mechanics 578, 171--222.

\bibitem[{De~Lange and De~Bruin(1998)}]{de1997sheet}
De~Lange, D.~F., De~Bruin, G.~J., 1998. {Sheet Cavitation and Cloud Cavitation,
  Re-Entrant Jet and Three-Dimensionality }. In: Biesheuvel, A., van Heijst,
  G.~F. (Eds.), Fascination of Fluid Dynamics. Springer, pp. 91--114.

\bibitem[{Edelbauer(2017)}]{Edelbauer:2017fx}
Edelbauer, W., 2017. Numerical simulation of cavitating injector flow and
  liquid spray break-up by combination of eulerian--eulerian and
  volume-of-fluid methods. Computers \& Fluids 144, 19--33.

\bibitem[{Egerer et~al.(2014)Egerer, Hickel, Schmidt, and
  Adams}]{Egerer:2014wu}
Egerer, C.~P., Hickel, S., Schmidt, S.~J., Adams, N.~A., 2014. Large-eddy
  simulation of turbulent cavitating flow in a micro channel. Physics of Fluids
  26~(8), 085102.

\bibitem[{Egerer et~al.(2016)Egerer, Schmidt, Hickel, and
  Adams}]{Egerer:2016it}
Egerer, C.~P., Schmidt, S.~J., Hickel, S., Adams, N.~A., 2016. Efficient
  implicit les method for the simulation of turbulent cavitating flows. Journal
  of Computational Physics 316~(C), 453--469.

\bibitem[{Franc and Michel(2005)}]{franc2004fundamentals}
Franc, J.-P., Michel, J.-M., 2005. Fundamentals of cavitation. Dordrecht:
  Springer science \& Business media.

\bibitem[{Furness and Hutton(1975)}]{Furness:1975wq}
Furness, R., Hutton, S., 1975. Experimental and theoretical studies of
  two-dimensional fixed-type cavities. Journal of Fluids Engineering 97~(4),
  515--521.

\bibitem[{Ganesh(2015)}]{Ganesh:2015uj}
Ganesh, H., 2015. Bubbly shock propagation as a cause of sheet to cloud
  transition of partial cavitation and  stationary cavitation bubbles forming
  on a delta wing vortex. Ph.D. thesis.

\bibitem[{Ganesh et~al.(2016)Ganesh, M{\"a}kiharju, and
  Ceccio}]{ganesh2016bubbly}
Ganesh, H., M{\"a}kiharju, S.~A., Ceccio, S.~L., 2016. Bubbly shock propagation
  as a mechanism for sheet-to-cloud transition of partial cavities. Journal of
  Fluid Mechanics 802, 37--78.

\bibitem[{Gnanaskandan and Mahesh(2016)}]{Gnanaskandan:2016bi}
Gnanaskandan, A., Mahesh, K., 2016. Large eddy simulation of the transition
  from sheet to cloud cavitation over a wedge. International Journal of
  Multiphase Flow 83, 86--102.

\bibitem[{Gopalan and Katz(2000)}]{gopalan2000flow}
Gopalan, S., Katz, J., 2000. Flow structure and modeling issues in the closure
  region of attached cavitation. Physics of Fluids 12~(4), 895--911.

\bibitem[{Hayashi and Sato(2014)}]{Hayashi:2014ik}
Hayashi, S., Sato, K., 2014. {Unsteady Behavior of Cavitating Waterjet in an
  Axisymmetric Convergent-Divergent Nozzle: High Speed Observation and Image
  Analysis Based on Frame Difference Method}. Journal of Flow Control,
  Measurement \& Visualization 02~(03), 94--104.

\bibitem[{He et~al.(2016)He, Chen, Leng, Wang, and Guo}]{He:2016bx}
He, Z., Chen, Y., Leng, X., Wang, Q., Guo, G., 2016. {Experimental
  visualization and LES investigations on cloud cavitation shedding in a
  rectangular nozzle orifice}. International Communications in Heat and Mass
  Transfer 76~(C), 108--116.

\bibitem[{Hickel et~al.(2006)Hickel, Adams, and
  Domaradzki}]{hickel2006adaptive}
Hickel, S., Adams, N.~A., Domaradzki, J.~A., 2006. An adaptive local
  deconvolution method for implicit les. Journal of Computational Physics
  213~(1), 413--436.

\bibitem[{Hickel et~al.(2014)Hickel, Egerer, and Larsson}]{hickel2014subgrid}
Hickel, S., Egerer, C.~P., Larsson, J., 2014. Subgrid-scale modeling for
  implicit large eddy simulation of compressible flows and shock-turbulence
  interaction. Physics of Fluids 26~(10), 106101.

\bibitem[{Jahangir et~al.(2018)Jahangir, Hogendoorn, and
  Poelma}]{Jahangir:2018id}
Jahangir, S., Hogendoorn, W., Poelma, C., 2018. Dynamics of partial cavitation
  in an axisymmetric converging-diverging nozzle. International Journal of
  Multiphase Flow 106, 34--45.

\bibitem[{Jakobsen(1964)}]{jakobsen1964mechanism}
Jakobsen, J., 1964. On the mechanism of head breakdown in cavitating inducers.
  Journal of Basic Engineering 86~(2), 291--305.

\bibitem[{Ji et~al.(2015)Ji, Luo, Arndt, Peng, and Wu}]{ji2015large}
Ji, B., Luo, X., Arndt, R.~E., Peng, X., Wu, Y., 2015. Large eddy simulation
  and theoretical investigations of the transient cavitating vortical flow
  structure around a naca66 hydrofoil. International Journal of Multiphase Flow
  68, 121--134.

\bibitem[{Ji et~al.(2014)Ji, Luo, Arndt, and Wu}]{ji2014numerical}
Ji, B., Luo, X., Arndt, R.~E., Wu, Y., 2014. Numerical simulation of three
  dimensional cavitation shedding dynamics with special emphasis on
  cavitation--vortex interaction. Ocean Engineering 87, 64--77.

\bibitem[{Kawanami et~al.(1997)Kawanami, Kato, Yamaguchi, Tanimura, and
  Tagaya}]{kawanami1997mechanism}
Kawanami, Y., Kato, H., Yamaguchi, H., Tanimura, M., Tagaya, Y., 1997.
  Mechanism and control of cloud cavitation. Journal of Fluids Engineering
  119~(4), 788--794.

\bibitem[{Koren(1993)}]{Koren:1993}
Koren, B., 1993. A robust upwind discretization method for advection, diffusion
  and source terms. Centrum voor Wiskunde en Informatica Amsterdam.

\bibitem[{Koukouvinis et~al.(2017)Koukouvinis, Naseri, and
  Gavaises}]{Koukouvinis:2016boa}
Koukouvinis, P., Naseri, H., Gavaises, M., 2017. Performance of turbulence and
  cavitation models in prediction of incipient and developed cavitation.
  International Journal of Engine Research 18~(4), 333--350.

\bibitem[{Kubota et~al.(1992)Kubota, Kato, and Yamaguchi}]{Kubota:1992eo}
Kubota, A., Kato, H., Yamaguchi, H., 1992. A new modelling of cavitating flows:
  a numerical study of unsteady cavitation on a hydrofoil section. Journal of
  Fluid Mechanics 240, 59--96.

\bibitem[{Kubota et~al.(1989)Kubota, Kato, Yamaguchi, and
  Maeda}]{Kubota:1989unsteady}
Kubota, A., Kato, H., Yamaguchi, H., Maeda, M., 1989. Unsteady structure
  measurement of cloud cavitation on a foil section using conditional sampling
  technique. Journal of Fluids Engineering 111~(2), 204--210.

\bibitem[{Laberteaux and Ceccio(2001)}]{laberteaux2001partial}
Laberteaux, K., Ceccio, S., 2001. Partial cavity flows. part 1. cavities
  forming on models without spanwise variation. Journal of Fluid Mechanics 431,
  1--41.

\bibitem[{Le et~al.(1993)Le, Franc, and Michel}]{Le:1993partial}
Le, Q., Franc, J.-P., Michel, J.-M., 1993. Partial cavities: global behavior
  and mean pressure distribution. Journal of Fluids Engineering 115~(2),
  243--248.

\bibitem[{Leroux et~al.(2004)Leroux, Astolfi, and Billard}]{Leroux:2004jl}
Leroux, J.~B., Astolfi, J.~A., Billard, J.~Y., 2004. An experimental study of
  unsteady partial cavitation. Journal of Fluids Engineering 126, 94--101.

\bibitem[{Leroux et~al.(2005)Leroux, Coutier-Delgosha, and
  Astolfi}]{Leroux:2005ii}
Leroux, J.-B., Coutier-Delgosha, O., Astolfi, J.~A., 2005. {A joint
  experimental and numerical study of mechanisms associated to instability of
  partial cavitation on two-dimensional hydrofoil}. Physics of Fluids 17~(5),
  052101--21.

\bibitem[{Lush and Skipp(1986)}]{lush1986high}
Lush, P., Skipp, S., 1986. High speed cine observations of cavitating flow in a
  duct. International Journal of Heat and Fluid Flow 7~(4), 283--290.

\bibitem[{Mauger et~al.(2012)Mauger, M{\'e}{\`e}s, Michard, Azouzi, and
  Valette}]{mauger2012shadowgraph}
Mauger, C., M{\'e}{\`e}s, L., Michard, M., Azouzi, A., Valette, S., 2012.
  Shadowgraph, schlieren and interferometry in a 2d cavitating channel flow.
  Experiments in fluids 53~(6), 1895--1913.

\bibitem[{Mihatsch et~al.(2015)Mihatsch, Schmidt, and
  Adams}]{mihatsch2015cavitation}
Mihatsch, M.~S., Schmidt, S.~J., Adams, N.~A., 2015. Cavitation erosion
  prediction based on analysis of flow dynamics and impact load spectra.
  Physics of Fluids 27~(10), 103302.

\bibitem[{Nurick(1976)}]{nurick1976orifice}
Nurick, W., 1976. Orifice cavitation and its effect on spray mixing. Journal of
  Fluids Engineering 98~(4), 681--687.

\bibitem[{{\"O}rley et~al.(2017){\"O}rley, Hickel, Schmidt, and
  Adams}]{Orley:2016db}
{\"O}rley, F., Hickel, S., Schmidt, S.~J., Adams, N.~A., 2017. Large-eddy
  simulation of turbulent, cavitating fuel flow inside a 9-hole diesel injector
  including needle movement. International Journal of Engine Research 18~(3),
  195--211.

\bibitem[{{\"O}rley et~al.(2015){\"O}rley, Trummler, Hickel, Mihatsch, Schmidt,
  and Adams}]{Orley:2015kt}
{\"O}rley, F., Trummler, T., Hickel, S., Mihatsch, M.~S., Schmidt, S.~J.,
  Adams, N.~A., 2015. {Large-eddy simulation of cavitating nozzle flow and
  primary jet break-up}. Physics of Fluids 27~(8), 086101--28.

\bibitem[{Payri et~al.(2004)Payri, Bermudez, Payri, and
  Salvador}]{payri2004influence}
Payri, F., Bermudez, V., Payri, R., Salvador, F., 2004. The influence of
  cavitation on the internal flow and the spray characteristics in diesel
  injection nozzles. Fuel 83~(4-5), 419--431.

\bibitem[{Petkov{\v{s}}ek and Dular(2013)}]{petkovvsek2013simultaneous}
Petkov{\v{s}}ek, M., Dular, M., 2013. Simultaneous observation of cavitation
  structures and cavitation erosion. Wear 300~(1-2), 55--64.

\bibitem[{Pham et~al.(1999)Pham, Larrarte, and Fruman}]{Pham:1999tp}
Pham, T., Larrarte, F., Fruman, D.~H., 1999. Investigation of unsteady sheet
  cavitation and cloud cavitation mechanisms. Journal of Fluids Engineering
  121~(2), 289--296.

\bibitem[{Reisman and Brennen(1996)}]{reisman1996pressure}
Reisman, G., Brennen, C., 1996. Pressure pulses generated by cloud cavitation.
  In: Proc. ASME Symp. on Cavitation and Gas - Liquid Flows in Fluid Machinery
  and Device FED 236. American Society of Mechanical Engineers, pp.
  319–--328.

\bibitem[{Reisman et~al.(1998)Reisman, Wang, and
  Brennen}]{reisman1998observations}
Reisman, G., Wang, Y.-C., Brennen, C.~E., 1998. Observations of shock waves in
  cloud cavitation. Journal of Fluid Mechanics 355, 255--283.

\bibitem[{Reitz and Bracco(1982)}]{reitz1982mechanism}
Reitz, R., Bracco, F., 1982. Mechanism of atomization of a liquid jet. The
  Physics of Fluids 25~(10), 1730--1742.

\bibitem[{Roe(1986)}]{Roe:1986}
Roe, P.~L., 1986. Characteristic-based schemes for the euler equations. Annual
  Review of Fluid Mechanics 18~(1), 337--365.

\bibitem[{Rudolf et~al.(2014)Rudolf, Hudec, Gr{\'\i}ger, and
  {\v{S}}tefan}]{rudolf2014characterization}
Rudolf, P., Hudec, M., Gr{\'\i}ger, M., {\v{S}}tefan, D., 2014.
  Characterization of the cavitating flow in converging-diverging nozzle based
  on experimental investigations. In: EPJ Web of conferences. Vol.~67. p.
  02101.

\bibitem[{Saito and Sato(2003)}]{Saito:2003us}
Saito, Y., Sato, K., 2003. Growth process to cloud-like cavitation on separated
  shear layer. In: ASME/JSME 2003 4th Joint Fluids Summer Engineering
  Conference. pp. 1379--1384.

\bibitem[{Sakoda et~al.(2001)Sakoda, Yakushiji, Maeda, and
  Yamaguchi}]{Sakoda:2001tn}
Sakoda, M., Yakushiji, R., Maeda, M., Yamaguchi, H., 2001. Mechanism of cloud
  cavitation generation on a 2-d hydrofoil. In: CAV 2001.

\bibitem[{Sato and Saito(2002)}]{Sato:2002vv}
Sato, K., Saito, Y., 2002. Unstable cavitation behavior in a
  circular-cylindrical orifice flow. JSME International Journal Series B Fluids
  and Thermal Engineering 45~(3), 638--645.

\bibitem[{Schmidt(2015)}]{Schmidt:2015wa}
Schmidt, S.~J., 2015. A low mach number consistent compressible approach for
  simulation of cavitating flows. Ph.D. thesis, Technical University of Munich,
  Technical University of Munich.

\bibitem[{Schmidt et~al.(2014)Schmidt, Mihatsch, Thalhamer, and
  Adams}]{schmidt2014assessment}
Schmidt, S.~J., Mihatsch, M.~S., Thalhamer, M., Adams, N.~A., 2014. {Assessment
  of erosion sensitive areas via compressible simulation of unsteady cavitating
  flows}. In: Kim, K.-H., Chahine, G., Franc, J.-P., Karimi, A. (Eds.),
  Advanced experimental and numerical techniques for cavitation erosion
  prediction. Springer, pp. 329--344.

\bibitem[{Schnerr et~al.(2008)Schnerr, Sezal, and Schmidt}]{Schnerr:2008jja}
Schnerr, G.~H., Sezal, I.~H., Schmidt, S.~J., 2008. Numerical investigation of
  three-dimensional cloud cavitation with special emphasis on collapse induced
  shock dynamics. Physics of Fluids 20~(4), 040703--10.

\bibitem[{Sou et~al.(2014)Sou, Bi{\c c}er, and Tomiyama}]{Sou:2014hja}
Sou, A., Bi{\c c}er, B., Tomiyama, A., 2014. {Numerical simulation of incipient
  cavitation flow in a nozzle of fuel injector}. Computers and Fluids 103~(C),
  42--48.

\bibitem[{Sou et~al.(2007)Sou, Hosokawa, and Tomiyama}]{Sou:2007jd}
Sou, A., Hosokawa, S., Tomiyama, A., 2007. {Effects of cavitation in a nozzle
  on liquid jet atomization}. International Journal of Heat and Mass Transfer
  50~(17-18), 3575--3582.

\bibitem[{Stanley et~al.(2011)Stanley, Barber, Milton, and
  Rosengarten}]{Stanley:2011gr}
Stanley, C., Barber, T., Milton, B., Rosengarten, G., Jun. 2011. {Periodic
  cavitation shedding in a cylindrical orifice}. Experiments in Fluids 51~(5),
  1189--1200.

\bibitem[{Stanley et~al.(2014)Stanley, Barber, and
  Rosengarten}]{Stanley:2014id}
Stanley, C., Barber, T., Rosengarten, G., 2014. Re-entrant jet mechanism for
  periodic cavitation shedding in a cylindrical orifice. International Journal
  of Heat and Fluid Flow 50~(C), 169--176.

\bibitem[{Stutz and Reboud(1997)}]{Stutz:1997ui}
Stutz, B., Reboud, J.~L., 1997. {Experiments on unsteady cavitation}.
  Experiments in Fluids 22~(3), 191--198.

\bibitem[{Sugimoto and Sato(2009)}]{Sugimoto:2009}
Sugimoto, Y., Sato, K., 2009. Visualization of unsteady behavior of cavitation
  in circular cylindrical orifice with abruptly expanding part. In: 13th
  International Topical Meeting on Nuclear Reactor Thermal Hydraulics. Vol.~13.
  pp. N13P1156, 1--10.

\bibitem[{Trummler et~al.(2018)Trummler, Rahn, Schmidt, and
  Adams}]{Trummler:2018AAS}
Trummler, T., Rahn, D., Schmidt, S.~J., Adams, N.~A., 2018. Large eddy
  simulations of cavitating flow in a step nozzle with injection into gas.
  Atomization and Sprays 28~(10), 931--955.

\bibitem[{Wade and Acosta(1966)}]{wade1966experimental}
Wade, R., Acosta, A., 1966. Experimental observations on the flow past a
  plano-convex hydrofoil. Journal of Basic Engineering 88~(1), 273--282.

\bibitem[{Wang et~al.(2017)Wang, Huang, Wang, Zhang, and Ding}]{Wang:2017hf}
Wang, C., Huang, B., Wang, G., Zhang, M., Ding, N., 2017. {Unsteady pressure
  fluctuation characteristics in the process of breakup and shedding of
  sheet/cloud cavitation}. International Journal of Heat and Mass Transfer 114,
  769--785.

\bibitem[{Wu et~al.(2017)Wu, Maheux, and Chahine}]{Wu:2017cda}
Wu, X., Maheux, E., Chahine, G.~L., 2017. {An experimental study of sheet to
  cloud cavitation}. Experimental Thermal and Fluid Science 83, 129--140.

\end{thebibliography}
\bibliographystyle{elsarticle-harv}
\biboptions{authoryear}

\end{document}